\documentclass[prd, 10pt,aps,nofootinbib, floatfix, notitlepage,twocolumn,longbibliography]{revtex4-2}

\usepackage{amsmath,amsfonts,amssymb,enumitem,fontawesome5}
\usepackage{mathrsfs}
\usepackage{multirow}
\usepackage{graphicx}
\usepackage[english]{babel} 
\usepackage{url} 
\usepackage{color,xcolor}
\usepackage{bbold}
\usepackage{adjustbox}
\usepackage[utf8]{inputenc} 
\usepackage[colorlinks=true,citecolor=blue,urlcolor=blue]{hyperref}
\usepackage[utf8]{inputenc}

\newcommand{\LCDM}{$\Lambda$CDM}

\begin{document}

\title{The Cosmology of Dark Energy Radiation}

\author{Kim V. Berghaus$^{1,2}$}
\author{Tanvi Karwal$^{3,4}$ }
\author{Vivian Miranda$^{2}$}
\author{Thejs Brinckmann$^{5,6}$}

\affiliation{$^1$Walter Burke Institute for Theoretical Physics, California Institute of Technology, CA 91125, USA}
\affiliation{$^2$C.N. Yang Institute for Theoretical Physics, Stony Brook University, NY 11794, USA}

\affiliation{$^3$Kavli Institute for Cosmological Physics, the University of Chicago, IL 60637, USA}
\affiliation{$^4$Center for Particle Cosmology, Department of Physics and Astronomy, University of Pennsylvania, Philadelphia, PA 19104, USA}

\affiliation{$^5$Dipartimento di Fisica e Scienze della Terra, Universit\`a degli Studi di Ferrara, via Giuseppe Saragat 1, I-44122 Ferrara, Italy}
\affiliation{$^6$Istituto Nazionale di Fisica Nucleare, Sezione di Ferrara, via Giuseppe Saragat 1, I-44122 Ferrara, Italy}

\begin{abstract}
In this work, we quantify the cosmological signatures of dark energy radiation -- a novel description of dark energy, which proposes that the dynamical component of dark energy is comprised of a thermal bath of relativistic particles sourced by thermal friction from a slowly rolling scalar field. 
For a minimal model with particle production emerging from first principles, we find that the abundance of radiation sourced by dark energy can be as large as $\Omega_{\text{DER}} = 0.03$, exceeding the bounds on relic dark radiation by three orders of magnitude. 
Although the background and perturbative evolution of dark energy radiation is distinct from Quintessence, we find that current and near-future cosmic microwave background and supernova data will not distinguish these models of dark energy. We also find that our constraints on all models are dominated by their impact on the expansion rate of the Universe.
Considering extensions that allow the dark radiation to populate neutrinos, axions, and dark photons, we evaluate the direct detection prospects of a thermal background comprised of these candidates consistent with cosmological constraints on dark energy radiation. 
Our study indicates that a resolution of $\sim 6 \, \text{meV}$ is required to achieve sensitivity to relativistic neutrinos compatible with dark energy radiation in a neutrino capture experiment on tritium. 
We also find that dark matter axion experiments lack sensitivity to a relativistic thermal axion background, even if enhanced by dark energy radiation, and dedicated search strategies are required to probe new parameter space. We derive constraints arising from a dark photon background from oscillations into visible photons, and find that viable parameter space can be explored with the LAte Dark Energy RAdiation (LADERA) experiment. 
\href{https://github.com/KBerghaus/class_der}{\faGithub} 
\end{abstract}

\maketitle

\date{\today}


\section{Introduction}
\label{Sec:Introduction}

Observations of type 1a supernovae first found evidence for the acceleration of the expansion of the Universe more than twenty years ago, driven by the mysterious dark energy \cite{SupernovaSearchTeam:1998fmf, SupernovaCosmologyProject:1998vns}. 
This result has been greatly strengthened since, with several hundred more measurements of type 1a supernovae \cite{Scolnic:2017caz, Brout:2022vxf} as well as other precision cosmological measurements of the cosmic microwave background (CMB) \cite{Sherwin:2011gv, Planck:2018vyg}, and baryon acoustic oscillations (BAO) \cite{Ross:2014qpa, Alam:2016hwk}, establishing that dark energy is the dominant component in the universe today, is approximately smooth, and constrained to be at most evolving slowly with a negative equation of state \cite{Planck:2018vyg}.

These properties have been deduced from the gravitational impact of dark energy on observables; no direct non-gravitational signal of dark energy has been measured to date. 
The lack of direct probes makes it challenging to obtain a fundamental understanding of what comprises dark energy. 
In the concordance flat $\Lambda$CDM model, dark energy is described simplistically, by a cosmological constant $\Lambda$ that does not evolve in time, with a constant equation of state of negative one. 
However, the fine-tuning of the value of $\Lambda$ is a notable shortcoming of this description. 
Addressing this shortcoming has proved challenging, but dark energy with a dynamical component is a promising avenue for alleviating this fine-tuning \cite{Abbott:1984qf, PhysRevLett.52.1461, Alberte:2016izw, PhysRevD.65.126003, Graham:2017hfr, Graham:2019bfu}. 

With this motivation, we study a novel dynamical dark energy model -- dark energy radiation, and explore its observables beyond just its gravitational impact with relevance for the plethora of precision data sets from experiments coming online within the next decade. 
Dynamical dark energy models beyond \LCDM\ are often explored via the phenomenological $w_0 w_a$ parameterization with two degrees of freedom, not motivated by an underlying theory. 
Slowly rolling scalar fields on the other hand are a minimal class of models that connect the properties of dark energy with particle physics, where traditionally the kinetic energy of the rolling scalar field makes up the dynamical component of the dark energy.

Coupling the scalar field to light degrees of freedom can modify this picture by giving rise to dissipation of the scalar field's energy into dark radiation, which can constitute the dominant dynamical component of the dark energy. 
This minimal extension, proposed in \cite{Berghaus:2020ekh}, has distinct predictions for the equation of state of dark energy and its time evolution, and additionally postulates the existence of a thermalized component of dark radiation, slowly growing as the universe expands, which we will refer to as dark energy radiation (DER). 
The energy density in DER can greatly exceed the energy density in the cosmic microwave background (CMB) as it only becomes important at late times and thereby avoids CMB limits on extra relativistic degrees of freedom. 
The characteristic temperature of DER is set by the energy scale of dark energy such that $T_{\text{DER}} \lesssim  \text{meV}$. 

In this work, we quantify the cosmological signatures of DER by implementing the model in the linear Boltzmann solver CLASS, and comparing it with precision cosmological data. 
We also evaluate the direct detection prospects of DER.
If components of DER couple to the Standard Model (SM), the isotropic thermal background with temperature $T_{\text{DER}}$ may be directly measurable by ongoing experimental programs. 
This is an exciting direction as the $\text{meV}$-scale thermal background is a smoking gun for DER, and would constitute direct non-gravitational evidence into the nature of dark energy. 
In particular, focusing on neutrinos, axions and dark photons, we recast the sensitivity of various dark matter direct-detection experiments to their relativistic counterparts with an energy density compatible with the  predictions of DER.

The outline of this paper is as follows. 
In Sec.~\ref{sec:thermal_friction}, we give an overview of the phenomenology of DER and discuss a microphysical particle model, minimal DER, where an axion couples to non-Abelian gauge fields, which realizes efficient particle production through dissipation from first principles.
In Sec.~\ref{sec:cosmology}, we explore the cosmological signatures of DER, focusing on minimal DER with temperature-dependent dissipation  and a toy model with constant dissipation, toy DER.
We include the scenarios of no dissipation, as well as the phenomenological $w_0w_a$-parameterization for dark energy for comparison. 
We describe the background and perturbative evolutions of all included models in Sec.~\ref{subsec:models}, detail data and methodology in Sec.~\ref{subsec:data}, and show our results for current and forecasted datasets in Sec.~\ref{subsec:results}. 
In particular, we quantify the maximum energy density in DER allowed by current data. 
Then using the results of minimal DER as our benchmark, $\Omega^{\text{max}}_{m\text{DER}} = 0.03$, we evaluate the direct-detection prospects of DER in Sec.~\ref{sec:particle_stuff}, considering neutrinos (\ref{subsec:neutrinos}), axions (\ref{subsec:axions}), and dark photons (\ref{subsec:dphotons}). 
Finally, we summarize our main findings in Sec.~\ref{sec:conclusion} and conclude with future directions and outlook.

\section{Dark Energy Radiation }
\label{sec:thermal_friction}

A dynamical scalar field $\varphi$, initially displaced from its equilibrium value and slowly evolving towards it is often invoked to model dark energy 
\cite{PhysRevD.37.3406,Wetterich:1987fm,Caldwell:1997ii}. 
Including particle production due to coupling $\mathcal{J_{\text{int}}}$ to a light sector, $\mathcal{L}_{\text{int}} = -\varphi \mathcal{J}_{\text{int}} \, $  can modify the dynamics and the predicted cosmological observables for dark energy. 
Many well-motivated models give rise to efficient particle production through non-thermal \cite{osti_957435,Anber:2009ua,Fonseca:2019ypl,DallAgata:2019yrr,Alexander:2022own} and thermal effects \cite{Bastero-Gil:2016qru,Berghaus:2019whh,Berghaus:2020ekh}. 
Here, we focus on the cosmology of particle production through thermal friction, which results in a friction term linear in $ \dot{\varphi}$ in the equation of motion of the scalar field \cite{Berera:1995ie,Berera:1995wh,Berera:1999ws,Berera:1998px, Berera:2008ar,Bastero-Gil:2010dgy,Bastero-Gil:2016qru,Laine:2016hma,Berghaus:2019whh,Agrawal:2022yvu}. 
Such a term can emerge from a Lagrangian description when the particles in the light sector $\mathcal{J}_{\text{int}}$ thermalize efficiently and constitute a  dark radiation component 
with a well-defined temperature 
$\rho_{\text{DER}} \equiv \frac{\pi^2}{30}  g_{*} T^4_{\text{DER}}$, 
where $g_*$ denotes the degrees of freedom in the light sector. 
Formally, $\Upsilon$ is determined by a Minkowskian correlator 
$\int d^4 \chi \langle \frac{1}{2} {\mathcal J}_{\text{int}}(\chi), \mathcal{J}_{\text{int}}(0)  \rangle$ \cite{Laine:2016hma}, 
where the microphysics of the light sector determines the temperature dependence of the friction coefficient 
$\Upsilon(\rho_{\text{DER}}) \equiv c_n \rho_{\text{DER}}^{\frac{n}{4}}$. 
Several explicit models have shown that it is possible to obtain friction coefficients with linear \cite{Bastero-Gil:2016qru,Berghaus:2020ekh}, and cubic \cite{Berera:2008ar,Bastero-Gil:2010dgy,Berghaus:2019whh,Berghaus:2020ekh} temperature dependencies where the dark radiation can be comprised of non-Abelian gauge bosons, dark photons, fermions, and axions.    
Such a thermal friction term has two interesting consequences for the evolution of the scalar field and its accompanying dark radiation.
On one hand, the linear friction term modifies the time-evolution of the scalar field. 
On the other hand, the scalar field sources dark radiation that doesn't redshift as expected for radiation since it gets replenished by the scalar field. 
The coupled equations of scalar field and dark radiation as a function of time $t$ evolve as
\begin{eqnarray} \label{eom}
    \ddot{\varphi} +\left(3H + \Upsilon \right) \dot{\varphi} +\frac{dV}{d\varphi}(\varphi) &=& 0 \,
    \nonumber \\ 
    \dot{\rho}_{\text{DER}} +4H \rho_{\text{DER}}  &=& \Upsilon\dot{\varphi}^2   \, ,
    \label{eq:eqs_of_motion}
\end{eqnarray}
where $V(\varphi)$ is the potential of the scalar field, $H =\dot{a}/a$ is the Hubble rate, and dots denote derivatives with respect to time $t$. In the regime in which $V(\varphi) \gg \rho_{\text{DER}}, \frac{1}{2} \dot{\varphi}^2$ \,, the sum of the scalar field and dark radiation energy densities is compatible with the observed properties of dark energy.

\subsection{Minimal dark energy radiation}
\label{subsec:mDER}

A minimal microphysical model for DER involves \cite{Berghaus:2020ekh} an axion-like field $\varphi$ coupled to a non-Abelian gauge group SU($N_c$): 
\begin{equation} 
    \label{eq:lagrangian}
    \mathcal{L} \propto \frac{\alpha_D}{16 \pi} \frac{\varphi}{f} G^a_{\mu \nu} \tilde{G}^{a\mu\nu} \,,
\end{equation}
where $G^a_{\mu \nu}$ ($\tilde{G^{a \mu \nu}} \equiv \epsilon^{\mu \nu \alpha \beta} G^a_{\alpha \beta}$ ) is the field strength of the gauge bosons of the non-Abelian group, $\alpha_D$ is the fine structure constant of the non-Abelian group, and $f$ is the symmetry breaking scale above which the effective Lagrangian in \eqref{eq:lagrangian} breaks down. A compelling feature of this setup is that an axion-like particle is protected by its symmetry from quantum and thermal mass corrections that could spoil the flatness of the the potential $V(\varphi)$.  
Nevertheless, in a thermal environment the operator $\langle \frac{\alpha}{16 \pi } G^a_{\mu \nu} \tilde{G}^{a\mu\nu} \rangle$ develops a non-zero expectation value that can be calculated in linear response theory whose dominant term is proportional to $\dot{\varphi}$. 

The macroscopic friction coefficient obtained from the proportionality constant is given by \cite{Berghaus:2019whh}
\begin{equation} \label{eq:friction}
    \Upsilon(\rho_{\text{DER}}) = \kappa(N_c) \alpha^5 \frac{\rho^\frac{3}{4}_{\text{DER}}  }{g_{*}^{\frac{3}{4}}f^2}  \equiv c_3 \rho^{\frac{3}{4}}_{\text{DER}}
\end{equation}
where  $\kappa(N_c)$ is an $O(10)$ number for $N_c = 2$ that can be extracted from the sphaleron rate \cite{Moore:2010jd} for general $N_c$. 
The source of the friction is due to non-perturbative sphalerons.\footnote{
The generation of a thermal mass that backreacts onto the potential $V(\varphi)$ is suppressed by the approximate shift symmetry of the axion, making this an ideal model to generate thermal friction.
} 
The non-Abelian gauge fields have non-trivial field configurations, that provide a barrier between vacuaa of the theory with differing topological Chern-Simons number $N_{CS}$. 
A nonzero speed of the scalar field ($ \langle \dot{\varphi} \rangle \neq 0$) due to an external potential $V(\varphi)$ biases transitions between differing vacuaa in one direction, thus generating sphalerons which decay into gauge bosons that comprise the radiation $\rho_{\text{DER}}$.  
If fermions are present in the dark radiation they can suppress or even shut off the friction as sphaleron transitions can build up chiral charge which counteracts the bias generated by $ \langle \dot{\varphi} \rangle \neq 0$. 
However, non-zero fermion masses violate chirality and allow a depletion of the chiral charge. 
If that happens fast enough, which it does for fermion masses that are larger than O($\alpha^2 T_{\text{DER}}$), then the friction is the same as in Eq.~\eqref{eq:friction}. 
For an in-depth discussion refer to \cite{Berghaus:2019whh, Berghaus:2020ekh}. 

In minimal DER, the dark radiation consists of dark non-Abelian gauge bosons and the axion. 
The gauge bosons and the axion are thermalized\footnote{
This is trivially satisfied if $\Upsilon > H$, since the interaction rate $\Gamma_{gg} \sim \alpha^2 T_{\text{DER}}$, and $\Gamma_{g\varphi} \sim \alpha^3 \frac{T_{\text{DER}}^3}{f^2}$ are also larger than $H$.
} 
and at a temperature much above the gauge bosons' confinement scale. 
Fermions charged under the gauge group may be part of the dark radiation bath,
however their mass must be larger than $\sim \alpha^2 T_{\text{DER}}$ to deplete chiral charge and maintain sphaleron friction. 
Additional particles may become thermalized in the dark radiation such as dark photons or relativistic Standard Model neutrinos. 
We focus on these scenarios, as well as various constraints and direct detection prospects of these extensions in Sec. \ref{sec:particle_stuff}. 
The cosmological constraints in Sec. \ref{sec:cosmology} are not sensitive to the details of beyond-standard-model (BSM) extensions. 
In fact, they are also applicable to other microphysical models whose macroscopic friction coefficient $\Upsilon(\rho_{\text{DER}})$ has the same temperature dependence as the ones we consider in our analysis ($n = 0; \, n =3$).

\section{Cosmological Constraints}
\label{sec:cosmology}

The energy density $\rho_{r,0}$ of radiation in our universe today is tightly constrained by the CMB such that $\Omega_{\text{r}} \equiv \frac{\rho_r,0}{\rho_c}
\simeq \Omega_m / z_{\text{eq}} \simeq 0.000092$, 
where $\rho_c = 3H^2_0 M^2_{\text{pl}}$ is the critical energy density, and $z_{\text{eq}}$ is the redshift at matter-radiation equality.
However, DER is not subject to those early-universe constraints on extra-relativistic species. 
Its energy density has always been sub-dominant to that of the dark energy scalar field, which is frozen at early times. 
Hence, DER is negligible in the early universe and avoids early-universe constraints but grows in the late universe and can constitute the dominant radiation component in the Universe today. 

In this section, we characterize the cosmological imprints of DER. 
Performing a Markov chain Monte Carlo (MCMC) analysis, we quantify the amount $\Omega_{\text{DER}} \equiv \frac{\rho_{\text{DER,0}}}{\rho_c}$ of DER today that is compatible with cosmological precision data. We build upon this key result of our cosmological analysis in our evaluation of direct detection prospects in Sec.~\ref{sec:particle_stuff}.

Our analysis explores and compares four beyond-\LCDM\ models.
We consider minimal DER as described in Sec.~\ref{subsec:mDER} with thermal friction coefficient $\Upsilon = c_3 \rho_{\text{DER}}^{\frac{3}{4}}$, as well as a phenomenlogical DER toy model with constant thermal friction coefficient $\Upsilon = c_0$.
To maximize the amount of energy density in dark radiation, we focus on the regime in which $\Upsilon \gg H$. 
We also examine the usual Quintessence dark energy scalar field with only Hubble friction and no thermal friction coefficient. 
For the scalar field models, we adopt a linear potential $V(\varphi) =  -C \varphi$ \cite{Berghaus:2020ekh}. 
Lastly, we include the phenomenological $w_0 w_a$-description of the equation of state of dark energy. 

In our implementation, we enforce that all the dark energy today is comprised of the components of the considered model such that there is no additional cosmological constant $(\Omega_{\Lambda} = 0)$. 

\subsection{Models}
\label{subsec:models}

We compare all cosmologies to a standard flat \LCDM\ universe, varying the usual 6 base parameters: 
the amplitude $A_s$ of the primordial curvature power spectrum at $k=0.05 $Mpc$^{-1}$ as $10^9A_s$; 
the tilt $n_s$ of this spectrum; 
the angular size $\theta_*$ of the sound horizon ($\theta_{\rm MC}$ for CAMB runs); 
the physical density $\Omega_bh^2$ of baryons; 
the physical density $\Omega_ch^2$ of cold dark matter;
and the optical depth $\tau$ to reionization. 
These have broad, uninformative priors. 
For neutrinos, we adopt the standard description, with one massive ($m_{\nu} = 0.06$eV) and two massless neutrinos.

\subsubsection{Minimal DER}
\label{sssec:DER}

Using the thermal friction coefficient $\Upsilon = c_3 \rho_{\text{DER}}^{\frac{3}{4}}$ for minimal DER, we find the background evolution of the (pseudo)-scalar field $\varphi$ and $\rho_{\text{DER}}$ with respect to conformal time $\tau$ to be 
\begin{equation}
    \label{eq:EOM1}
    \varphi'' +\left(2aH +  a c_3 \rho_{\text{DER}}^{\frac{3}{4}} \right)\varphi' +a^2 C = 0 \,,
\end{equation}
\begin{equation}
    \label{eq:EOM2}
    \rho_{\text{DER}}' + 4 a H \rho_{\text{DER}}  = \frac{c_3 \rho_{\text{DER}}^{\frac{3}{4}}}{a} {\varphi'}^2 \,,
\end{equation}
where primes denote derivatives with respect to conformal time $ \tau$.
To be compatible with the properties of dark energy, the scalar field has to evolve slowly such that $V(\varphi) = - C \varphi$ is the dominant contribution to dark energy, a condition determined by the ratio of the slope $C$ of the potential, and the size of the friction $\Upsilon$.
At the background level, these two parameters are degenerate with one another. 
The equation of state $w \equiv \frac{p}{\rho}$ is determined by the weighted sum of the contributions of the equations of state of the scalar field and DER,
\begin{align}
    w^{\text{DER}}_{\text{de}} &=   
    \frac{\frac{ {\varphi'}^2}{2a^2} + C \varphi}{\frac{ {\varphi'}^2}{2a^2} - C \varphi}   \left(\frac{\rho_{\varphi}}{\rho_{\varphi} +\rho_{\text{DER}} } \right)
    +\frac{1}{3} \left(\frac{\rho_{\text{DER}}}{\rho_{\varphi} +\rho_{\text{DER}} } \right) \\ \nonumber
    & \approx - 1 \left(\frac{\rho_{\varphi}}{\rho_{\varphi} +\rho_{\text{DER}} } \right) + \frac{1}{3} \left(\frac{\rho_{\text{DER}}}{\rho_{\varphi} +\rho_{\text{DER}} } \right) 
\end{align}

On large angular scales ($\ell <100$), the cosmic microwave background (CMB) power spectrum is dominated by the late integrated Sachs-Wolfe effect (i.e. changes to the gravitational potential that CMB photons travel through), 
which is sensitive to spatial fluctuations of the dark energy density in the late universe.
These spatial fluctuations are governed by scalar field dynamics $\delta \varphi$, with  subleading contributions from DER $\delta_{\text{DER}} \equiv \frac{\delta \rho_{\text{DER}}}{\rho_{\text{DER}}}$.   

The evolution of the linear perturbations of the scalar field $\delta \varphi$, and DER $\delta_{\text{DER}}$ in Fourier space, are determined by the coupled set of differential equations

\begin{align} \label{eq:pert1}
    & \delta \varphi'' 
    + \left(2aH + a c_3 \rho_{\text{DER}}^{\frac{3}{4}} \right)\delta \varphi' 
    + k^2 \delta \varphi \\ \nonumber
    &= -\frac{\varphi'}{2} 
    \left(h' 
    + \frac{3}{2}a c_3 \rho_{\text{DER}}^{\frac{3}{4}} \delta_{\text{DER}} \right) \, ,
\end{align}
\begin{align} \label{eq:pert2}
    \delta_{\text{DER}}' 
    =& -2\frac{h'}{3} 
    - \frac{4}{3}\theta_{\text{DER}} 
    + 2\frac{c_3 \rho_{\text{DER}}^{\frac{3}{4}}}{a \rho_{\text{DER}}} \delta \varphi' \varphi' \\ 
    & \nonumber
    -\frac{1}{4} \frac{c_3 \rho_{\text{DER}}^{\frac{3}{4}} {\varphi'}^2}{a \rho_{\text{DER}}} \delta_{\text{DER}} \, ,
\end{align}
\begin{equation}
    \theta_{\text{DER}}' 
    = \frac{k^2}{4} \delta_{\text{DER}}  
    +k^2 \frac{3 c_3 \rho_{\text{DER}}^{\frac{3}{4}}}{4 a \rho_{\text{DER}}} \varphi' \delta \varphi 
    - \frac{c_3 \rho_{\text{DER}}^{\frac{3}{4}}}{a \rho_{\text{DER}}} {\varphi'}^2 \theta_{\text{DER}} \, .
\end{equation}
where $h$ denotes the metric perturbation in synchronous gauge
and $\theta_{\text{DER}} \equiv i k^i v_i$ denotes the DER velocity perturbation.\footnote{Higher order moments are suppressed by thermalization.} 

As usual in scalar-field dark energy models,  the perturbations of the scalar field have a source term $-\frac{h'\varphi'}{2}$ that couples the dark energy perturbations to matter fluctuations via gravity.  
However, the DER model has an additional source term that couples the scalar field perturbations to the DER perturbations via the thermal friction $\Upsilon$. 
In minimal DER, the friction and hence this term depend on the dark radiation density $\rho_{\rm DER}$. 

The inhomogeneous energy flux from the scalar field to DER can dominate observables by giving rise to a growing mode that enhances the dark energy fluctuations beyond linear growth in parts of the parameter space in which  $\Upsilon$ exceeds $H$ by many orders of magnitude, such that $\Upsilon \gg H$ and $\Upsilon/3H \gg k/aH$, for modes observable in the CMB $k> 0.002 \, \text{Mpc}^{-1}$. 
 When rewriting Eq.~\eqref{eq:pert1} and Eq.\eqref{eq:pert2} in terms of the variable $z = k/aH$, one can show that fluctuations grow as $\delta \varphi \sim (aH/k)^9$ in this regime, which has been explored in detail in the context of warm inflation \cite{Chris_Graham_2009, Bastero-Gil:2011rva}. For our purpose of studying the phenomenology of DER, we solve the full system of equations in CLASS, and thus do not have to rely on analytical arguments.

We find that in the described regime in which $\Upsilon$ exceeds $H$ by many orders of magnitude,  linear perturbation theory breaks down for modes on large scales $k\sim 0.002 \, \text{Mpc}^{-1}$, and the results we obtain calculating CMB multipoles using the linear Boltzmann solver CLASS are not valid. 
Implementing a rigorous nonlinear treatment of those fluctuations is beyond the scope of this analysis. However, we note that approximate smoothness is a key property characterizing dark energy and therefore the parts of parameter space predicting large fluctuations in our model are not compatible with observations of our Universe.
The nonlinear regime corresponds to the region of parameter space in which the scalar field has a steep potential and faces significant thermal friction. 
The strong friction slows the field, minimizing $\dot{\varphi}$ such that dark radiation is suppressed, and dark energy approaches an equation of state of $w = -1$. 
The perturbations in this regime are unstable, as a large $c_3$ amplifies the fluctuations in $\varphi'$ and $\delta_{\rm DER}$ as seen from Eq.~\eqref{eq:pert1}. 
This sensitivity of the perturbations to $c_3$ breaks the background-level degeneracy between the slope $C$ of the potential and the size $ \propto c_3$ of the friction and the viable parameter space is effectively pushed towards shallower potentials, where $\Upsilon$ only exceeds $H$ by a few orders of magnitude.  

We explore the regime mDER regime, varying both $C \in [10^{-8}, 10^{-6}]$ in units of $[M_{\text{pl}}\text{Mpc}^{-2}]$ 
and $\log_{10} c_3 \in [5.0,8.7]$ in units of $[ 3^{-\frac{3}{4}}M_{\text{pl}}^{-\frac{3}{2}} \text{Mpc}^{\frac{1}{2}}]$\footnote{
The unconventional units of $c_3$ presented above match the CLASS convention of units that our public code and analysis chains follow. 
For reference, for $\varphi$ in units of [eV] and $V(\varphi)$ in units of [eV$^4$], $c_3$ would have units of [eV$^{-2}$]. 
}, beyond the usual \LCDM\ parameters. 
In terms of the fundamental parameters of the microphysical theory, $c_3 = \frac{\alpha^5}{f^2} \left(\frac{30}{\pi^2 g_{*}} \right)^{\frac{3}{4}}$, such that for $g_* = 10$, and $\alpha = 0.1$, the range explored corresponds to values of $0.002 \, \text{GeV}  \leq f \leq  0.08 \, \text{GeV}$. There is a degeneracy between $\alpha$ and $f$, such that for $\alpha = 0.005$ the sampled range would correspond to $1 \, \text{keV}  \leq f \leq  45 \, \text{keV}$ instead.
This combination of parameters allows for up to O(10)$\%$ of $\Omega_{\text{DER}}$.

\subsubsection{Toy DER}

In the toy model implementation of DER we treat the thermal friction coefficient $\Upsilon = c_0$ as constant. 
The background equations are then given by 
\begin{equation}
    \label{eq:EOM1}
    \varphi'' +\left(2aH + a c_0 \right) \varphi' +a^2 C = 0 \, , {\rm and}
\end{equation}
\begin{equation}
    \label{eq:EOM2}
    \rho_{\text{DER}}' + 4 a H \rho_{\text{DER}}  = \frac{c_0}{a} {\varphi'}^2 \,,
\end{equation}
and the fluctuations simplify to
\begin{equation}
    \delta \varphi'' + \left(2aH +  a c_0\right) \delta  \varphi' + k^2 \delta \varphi = -\frac{\varphi'}{2} h' \, ,
\end{equation}
\begin{align}
    \delta_{\text{DER}}' 
    =& -2\frac{h'}{3} 
    - \frac{4}{3}\theta_{\text{DER}} 
    + 2\frac{c_0} {a \rho_{\text{DER}}} \delta \varphi' \varphi' 
    - \frac{c_0 {\varphi'}^2}{a \rho_{\text{DER}}} \delta_{\text{DER}} \, ,
\end{align}
\begin{equation}
    \theta_{\text{DER}}' 
    = \frac{k^2}{4} \delta_{\text{DER}}  
    +k^2 \frac{3 c_0} {4 a \rho_{\text{DER}}} \varphi' \delta \varphi 
    - \frac{c_0} {a \rho_{\text{DER}}} {\varphi'}^2 \theta_{\text{DER}} \, .
\end{equation}

In contrast to minimal DER, in the case of constant friction there is no growing mode, and the fluctuations of the scalar field remain decoupled from the radiation perturbation. 
These equations remain perturbative for all parts of parameter space. 
The constraints of the toy model are dominated by observables sensitive to the background evolution of the universe $H(z)$. 

In exploring toy DER, we fix $C = 1\times 10^{-6}M_{\text{pl}}\text{Mpc}^{-2}$, and vary the friction coefficient $c_0$ in units of [$\text{Mpc}^{-1}$] as $\log_{10}c_0 \in [-1,1]$, 
having confirmed that our results are robust to simultaneously varying the slope $C$ and the friction parameter $c_0$. The lower prior limit of $c_0$ allows for $\Omega_{\text{DER}} \approx 0.1$, and the upper limit corresponds to the asymptotic limit of a frozen scalar field with $\Omega_{\text{DER}} \approx 0$.

\subsubsection{Quintessence}
\label{sssec:Quintessence}

The evolution of the Quintessence scalar field $\varphi_Q$  is dominated by Hubble friction.
The equations are
\begin{equation}
    \varphi_Q'' +2aH  \varphi_Q' +a^2 C = 0      \, , \, {\rm and}
\end{equation}
\begin{equation}
    \delta \varphi_Q'' + 2aH +  \delta \varphi_Q' + k^2 \delta \varphi_Q = -\frac{h \varphi_Q'}{2} \, .
\end{equation}
The equation of state of dark energy is then given by
\begin{equation}
    w^Q_{\text{de}} 
    = \frac{\frac{ {\varphi_Q'}^2}{2a^2} 
    + C \varphi_Q}{\frac{ {\varphi_Q'}^2}{2a^2} 
    - C \varphi_Q}  
    \approx -1  
    + \frac{\frac{ {\varphi_Q'}^2}{2a^2}}{- C \varphi_Q }\, .
\end{equation}

The fluctuations in the Quintessence model do not currently contribute observable signatures, and its constraints are dominated by the background evolution of dark energy. 

We explore Quintessence by varying the linear slope $C \in [5\times 10^{-9}, 2\times 10^{-7}]$ of the scalar potential in units of $[M_{\text{pl}}\text{Mpc}^{-2}]$. The steepest slope in this prior allows for up to $\Omega_{\text{scf,k}} \approx 0.1$, and the shallowest slope asymptotes to $\Omega_{\text{scf,k}} \approx 0$.

\subsubsection{$w_0w_a$-Parameterization}

The $w_0w_a$ model, also refered to as Chevallier-Polarski-Linder (CPL) model \cite{Chevallier:2000qy,Linder:2002et}, is defined by parameterizing the equation of state of dark energy as 
\begin{equation}
    w(a)  = w_0 + (1-a) w_a \,.
\end{equation}
It has been widely used to characterize a dynamical dark energy equation of state, and we use it as a comparison to the other models. 
We follow the default treatment of the perturbations of this model in CLASS and CAMB, detailed in \cite{Fang:2008sn, Ballesteros:2010ks}, noting that the constraints we find are dominated by the background dynamics. 

For $w_0w_a$, we vary the dimensionless equation of state parameters today $w_0$ and deep in the past $w_a$ as $w_0 \in [-1, -0.01]$ and $w_0+w_a \in [-1, -0.01]$.
These priors exclude the phantom regime $w(z) < -1$ from $w_0w_a$, allowing a direct comparison between beyond-\LCDM\ cosmologies as Quintessence and DERs do not cross this physical boundary.

\subsection{Data and Methodology} 
\label{subsec:data}

We explore the parameter space of these four descriptions of dark energy, along with standard \LCDM\ through Markov chain Monte Carlos (MCMCs) \cite{Lewis:2002ah,Lewis:2013hha}, run using Cobaya \cite{Torrado:2020dgo,2019ascl.soft10019T}.
Chains that attain a Gelman-Rubin criterion $R-1 < 0.02$ after removing 50\% as burn-in are considered converged. 
For minimal DER, toy DER, and Quintessence, we use a modified version of the Boltzmann code CLASS \cite{Blas:2011rf}, which we make publicly available along with our MCMC chains\footnote{
\href{https://github.com/KBerghaus/class_der}{\tt https://github.com/KBerghaus/class\_{der}}  }. 
For the $w_0 w_a$ parameterization and reference \LCDM\ models, we run the Boltzmann code CAMB \cite{Lewis:1999bs,Howlett:2012mh} with no additional changes. 

We use data sets concordant under the standard \LCDM\ model to explore these cosmologies, dividing results into two sections -- current data and forecasts. 
In currently available data, we fit to the following likelihoods: 
\begin{itemize}
    \item CMB temperature and polarization measurements by Planck \cite{Aghanim:2019ame}: low-$\ell$ TT and EE, and high-$\ell$ TTTEEE cut off at $\ell = 1300$\footnote{
    The cutoff $\ell \leq 1300$ for CMB data was chosen to minimize the impact of the non-linear evolution perturbations. 
    We reduced $\ell_{\rm cutoff}$ from $2500$ until the Planck $\chi^2$ difference between turning CMB non-linear lensing on and off in \LCDM\ was $\Delta \chi^2 \lesssim 1$. } 
    to remove scales impacted by non-linear signatures of the new cosmologies;
    \item baryon acoustic oscillation (BAO) measurements from SDSS MGS DR7 \cite{Ross:2014qpa} and DR12 \cite{Alam:2016hwk}; and 
    \item Pantheon supernovae (SNe) measurements \cite{Scolnic:2017caz}. 
    A more recent SNe data set, Pantheon+ has been released \cite{Brout:2022vxf}. However, we use the previous Pantheon data to be conservative as there is a significant shift in the value of $\Omega_m$ between the two iterations that is not well understood. 
\end{itemize}

For forecasts, we replace the Pantheon data with optimistic projections for measurements of supernovae by the Nancy Grace Roman Space Telescope, henceforth referred to as Roman, concentrating on the dataset that would be most sensitive to late-time background deviations from \LCDM. 
The CMB scales most impacted by such modified evolution of dark energy are $\ell < 50$, where error bars are limited by large cosmic variance.  
While BAOs are also expected to constrain late-time deviations of dark energy from \LCDM, we do not perform projections for improvements in these measurements, but these can be performed for example for the Dark Energy Spectroscopic Instrument (DESI). 
Considering the Allz case of \cite{Hounsell:2017ejq}, including modelling of all Roman systematics for $z<3$, an improvement of figure of merit of $\sim 2.3$ is expected on the $w_0-w_a$ plane relative to current CMB+BAO measurements. 
A similar improvement can be expected for Quintessence and DERs, as these models of dark energy are only expected to deviate from \LCDM\ at late times. 

We also replace the high-$\ell$ Planck data with forecasts for the Simons Observatory (SO)\footnote{\url{https://github.com/simonsobs/LAT_MFLike}}, again cutting at $\ell = 1300$ to remove the scales impacted by non-linear physics. 

Hence, our forecast data set uses:
\begin{itemize}
    \item CMB data from Planck low-$\ell$ TT and EE \cite{Aghanim:2019ame} and SO projections for high-$\ell$ TTTEEE cut at $\ell = 1300$;
    \item BAO data from SDSS MGS DR7 \cite{Ross:2014qpa} and SDSS DR12 \cite{Alam:2016hwk}; and
    \item optimistic Roman forecasts up to $z=3$ \cite{Hounsell:2017ejq}. 
\end{itemize}

It is important to note that the two projected forecasts -- Roman and SO use 
different fiducial cosmologies, however 
the Roman projections are
the dominant driving force for our constraints on late-time dark energy.

Finally, to directly compare the four dark energy cosmologies to \LCDM, we look at their posteriors on the normalization-adjusted expansion rate $E(z) \equiv H(z)/H_0$ and the equation of state $w(z)$ of dark energy as functions of redshift by post-processing chains. 
\begin{figure*}
    \centering
    \includegraphics[width=0.245\textwidth]{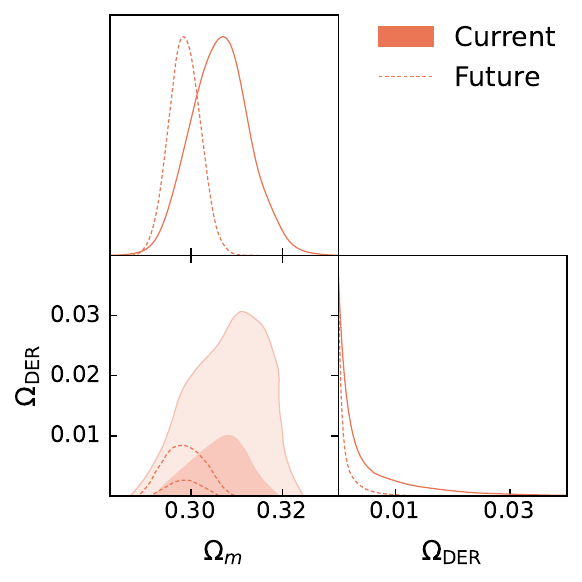}
    \includegraphics[width=0.245\textwidth]{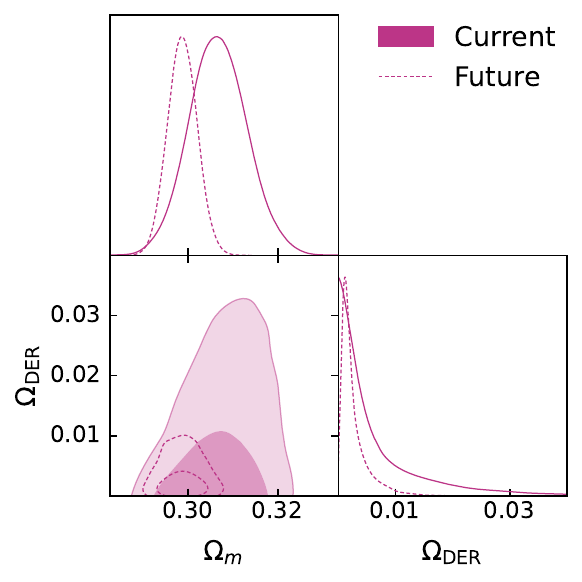}
    \includegraphics[width=0.245\textwidth]{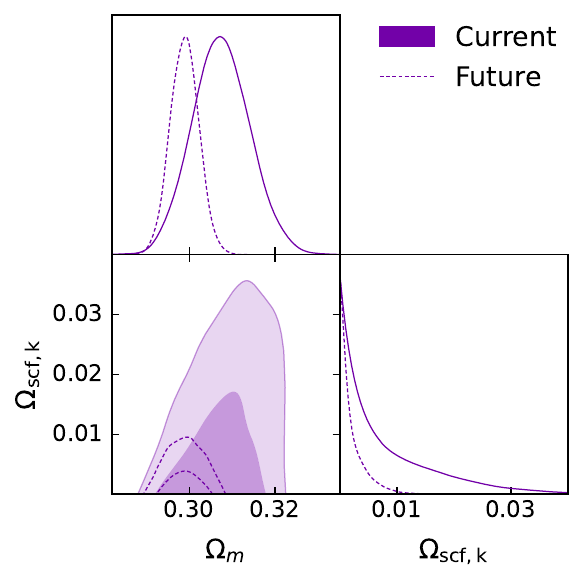}
    \includegraphics[width=0.245\textwidth]{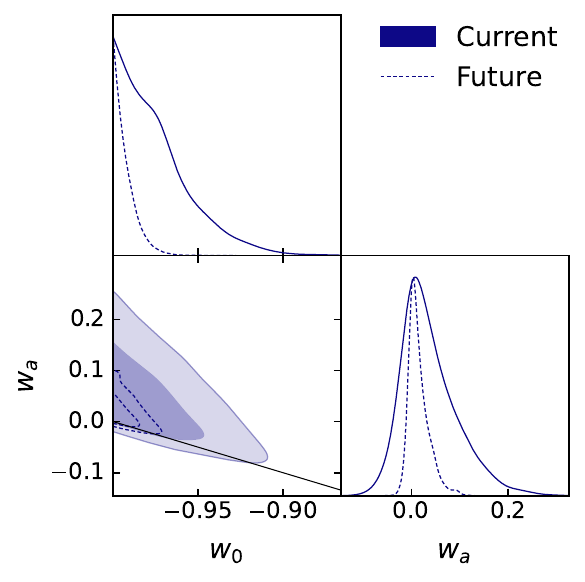}
    \caption{
    Constraints from our current and future data sets for the cosmologies we explore, showing just the dark-energy parameters that deviate from $\Lambda$. 
    In order from left to right, 
    (i) mDER and (ii) tDER parameterised by the fractional energy density $\Omega_{\rm DER}$ of DR, 
    (iii) a linear scalar field parameterised by the fractional energy density $\Omega_{\rm scf, k}$ in its kinetic energy and 
    (iv) phenomenological $w_0$ and $w_a$. 
    For $w_0w_a$, we restrict the priors to exclude the phantom regime for a more direct comparison between cosmologies, with the bounds shown by the black line. 
    }
    \label{fig:tri_now_fut_de_params}
\end{figure*}

First, we thin chains down to $O(2000)$ steps, verifying that the 1 and 2D posteriors are reproduced despite the thinning, adjusting the exact number of steps retained accordingly. 
Then, we re-run CLASS for each point, querying the expansion rate and the equation of state of the dark energy component over all cosmic time. 
The resultant array is interpolated and queried again for $H(z)$ and $w(z)$ at specific redshifts spanning the range $z \in [0,2]$. 
These are saved as additional derived parameters in the chains, and finally we use GetDist \cite{Lewis:2019xzd} to determine their 1D errors and means. 
A similar process also allows us to obtain $1\sigma$ error regions for the matter power spectrum $P(k)$ and the CMB temperature power spectrum $C_\ell^{TT}$ for each of the cosmologies investigated. 

\begin{figure}
    \centering
    \includegraphics[width=0.49\textwidth]{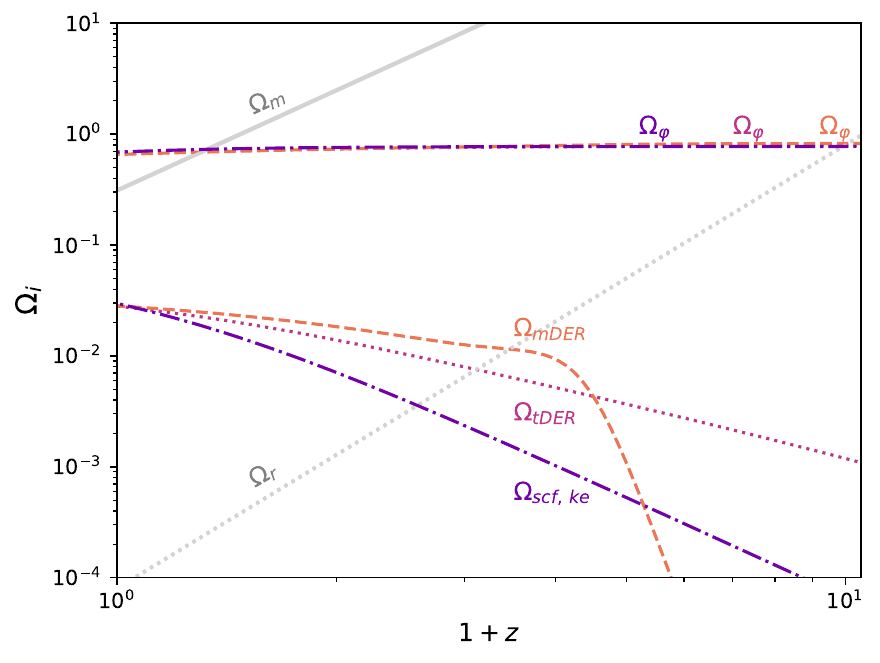}
    \caption{
    The fractional energy densities $\Omega_i \equiv \rho_i / \rho_{\rm total, \, 0}$ in various dark energy models relative to the total energy density $ \rho_{\rm total, \, 0}$ today. 
    Here $\Omega_\varphi$ is the total energy density in the scalar field in mDER (dashed, orange), tDER (dotted, pink) and Quintessence (dash-dotted, purple); 
    $\Omega_{\rm mDER}$ (dashed orange) and $\Omega_{\rm tDER}$ (dotted pink) are DER densities and $\Omega_{\rm scf,\, ke}$ (dash-dotted purple) is the kinetic energy of the Quintessence scalar. 
    For reference, we also show the energy densities in matter (solid grey) and standard radiation (dotted grey) for all three cosmologies, which visually overlap. 
    We choose a cosmology that has $\Omega_{\rm mDER} \simeq 3\%$ today, close to to its $2\sigma$ limit in Fig.~\ref{fig:tri_now_fut_de_params}. 
    We then choose $\Omega_{\rm tDER} \,,\, \Omega_{\rm scf,\, ke} \sim 3\%$ today to match. 
    }
    \label{fig:bg_omega_der}
\end{figure}

\subsection{Results} 
\label{subsec:results}

\begin{figure*}
    \centering
    \includegraphics[width=0.49\textwidth]{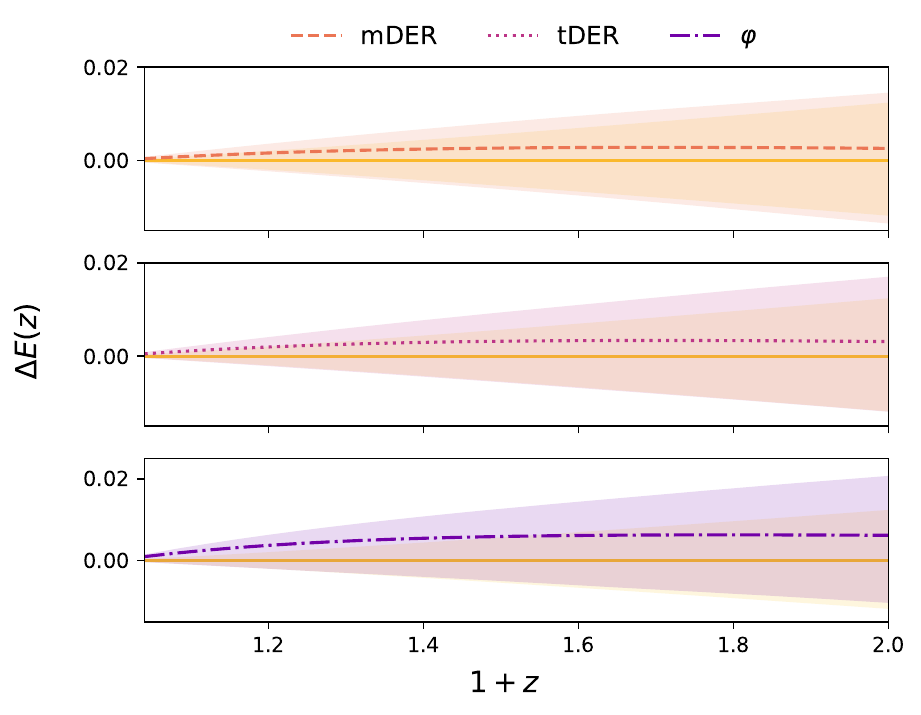}
    \includegraphics[width=0.49\textwidth]{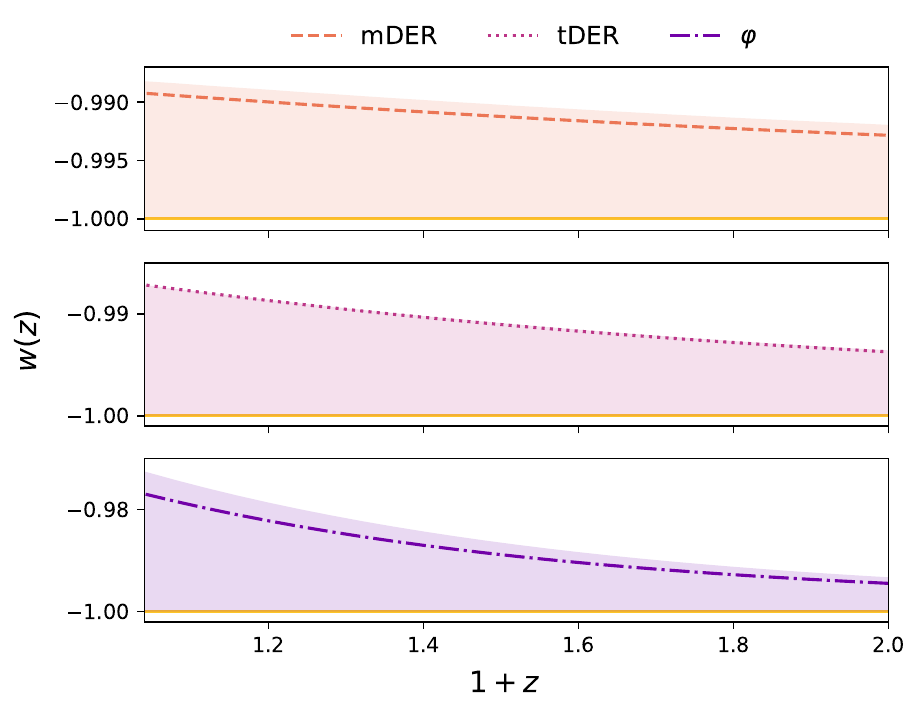}
    \caption{
    The expansion rate $E(z) \equiv H(z)/H_0$ (\textit{left}) and the equation of state $w(z)$ of dark energy (\textit{right}) in different cosmologies compared to those in \LCDM, based on constraints from current data including Planck temperature and polarization, BAO and Pantheon supernova data. 
    Curves correspond to the mean while shaded regions correspond to $1\sigma$ limits. 
    The solid curves and yellow regions correspond to \LCDM\ and are identical per column. 
    Top to bottom, we show minimal DER, toy DER, and a linear scalar field forming dark energy. 
    }
    \label{fig:E_w_of_z_now}
\end{figure*}

Constraints from our current and future data sets for the four dark energy cosmologies we explore are summarized in Fig.~\ref{fig:tri_now_fut_de_params} and Tables~\ref{tab:results_current} and \ref{tab:results_future}. 
In each case, we concentrate on the parameters that capture the divergence of dark energy from a cosmological constant. 
For DERs, we show the fraction $\Omega_{\rm DER}$ of DER today; for Quintessence, we show the fractional contribution $\Omega_{\rm scf,\,k}$ from the kinetic energy of the scalar $\varphi$; and for $w_0w_a$, constraints are represented by the current equation of state $w_0$ of dark energy and that $w_a$ deep in the past.
For DERs and Quintessence, we only find upper bounds on the allowed fractions. 

While $w_0w_a$ is also consistent with a cosmological constant, \LCDM\ lies at the edge of the prior region in $w_0 \in [-1, 0]$ and $(w_0 + w_a) \in [-1,0]$, such that derived parameters eg. $w_a, \, H(z)$, etc. suffer edge effects. 
The prior bound that excludes the phantom region in $w_0w_a$ is shown by the black line in the rightmost plot in Fig.~\ref{fig:tri_now_fut_de_params}, showing that \LCDM\ is within the $1\sigma$ bound. 
The extent of the contour beyond this bound is due to GetDist smoothing 2D contours. 

An immediate difference apparent when comparing the constraints from current data to forecasts is the shift in means. 
The means of the posteriors of forecasts are driven by the choice of fiducial cosmology, and do not reflect the true preference of future data. 
Moreover, small deviations from \LCDM\ in forecasts may be driven by the small differences in fiducial cosmologies for the CMB and SNe data. 
We hence comment only on the sensitivity of and improvement in parameter errors from future data, concentrating on the extent of posterior contours and not their locations in parameter space. 

For mDER  and tDER, we find that $\Omega_{\rm DER}$ can contribute upto $3\%$ of the energy budget of the Universe today at $2\sigma$, from the 2D marginalised posteriors in Fig.~\ref{fig:tri_now_fut_de_params}. 
These bounds are $\mathcal{O}(10^2)$ greater than the contribution $\Omega_r$ of traditional \LCDM\ radiation species today and exceed the maximum viable amount of dark radiation $\Omega^{\Delta N_{\text{eff}}<0.5}_{\text{dr}} \leq 6 \times 10^{-6}$ in the early universe \cite{Planck:2018vyg} by a factor of $5000$.
Forecasts, which assume a fiducial \LCDM\ universe, still allow $\Omega_{\rm DER} \simeq 1\%$, that is, even if the true underlying model of the Universe resembles \LCDM, DER can only be constrained to the percent-level with future CMB and supernova data, still exceeding the limits on cosmic dark radiation by three orders of magnitude. 

The DER explored here is able to evade these bounds as $\Omega_{\rm DER}$ grows in the late universe, being dynamically produced through a coupling to dark energy, while other radiation species are strongly constrained via their presence in the early universe, particularly through constraints on the time $z_{\rm eq}$ and scale $\ell_{\rm eq}$ of matter-radiation equality. 
While DER is also present in the early universe, its energy density is negligible at all times before $z \gtrsim 3$. 
This is visualized in Fig.~\ref{fig:bg_omega_der}, which shows the energy densities of mDER, tDER and Quintessence along with matter and radiation evolving with redshift. 
We arbitrarily choose to represent an mDER cosmology with $\Omega_{\rm mDER} = 2.8\% \simeq 2\sigma$ upper bound today, with equivalent $\Omega_{\rm tDER},\, \Omega_{\rm scf,\, k}$ today for tDER and Quintessence. 
Hence, the energy densities in the scalar potential $\Omega_\varphi$, matter $\Omega_m$ and radiation $\Omega_r$ roughly match and the main difference between the models at the background level becomes the evolution of DER or the scalar kinetic energy. 
For all three cosmologies, the dynamical subcomponent of dark energy is negligible in the early universe and grows with time. 
The sharpest growth is seen in mDER, where $\Omega_{\rm mDER}$ rapidly grows when the scalar field transitions from being dominated by Hubble friction to thermal friction, $\Upsilon(z) \sim H(z)$, then settles to a shallow growth. 

\begin{figure}
    \centering
    \includegraphics[width=0.49\textwidth]{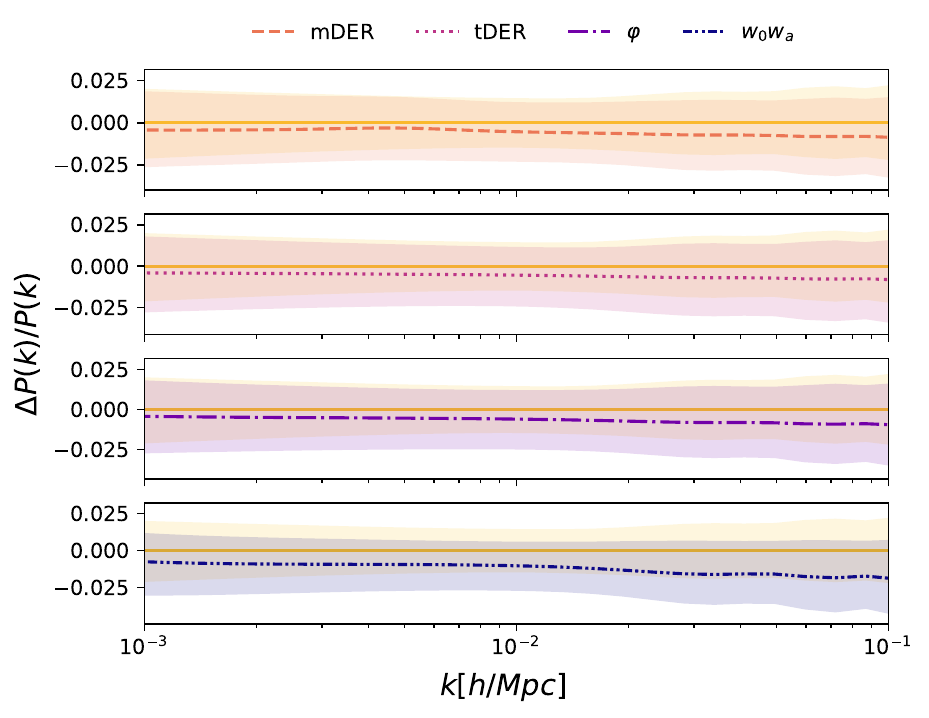}
    \caption{
    The matter power spectra $P(k)$ at linear scales for different cosmologies are shown relative to \LCDM, based on constraints from current data including Planck temperature and polarization, BAO and Pantheon supernova data. 
    Curves correspond to the mean while shaded regions correspond to $1\sigma$ limits. 
    The solid curves and yellow regions correspond to \LCDM\ and are identical. 
    Top to bottom, we show minimal DER, toy DER, a linear scalar field forming dark energy, and $w_0w_a$. 
    }
    \label{fig:P_of_k_now}
\end{figure}

Comparing constraints in Fig.~\ref{fig:tri_now_fut_de_params}, one may note the similarity between the posteriors for Quintessence kinetic energy $\Omega_{\rm scf,\, k}$ and DERs $\Omega_{\rm DER}$. 
We find that these constraints are dominated by the impact of dark energy on the background expansion of the Universe, specifically, supernova data. 
While their upper limits are similar, these species have distinct background evolution as shown in Fig.~\ref{fig:bg_omega_der}. 
The distinct slopes in the evolution of the dynamical component of dark energy translate to the small differences in their constraints. 
As the DER component in mDER has the shallowest slope at late times, the contribution of its dynamical component $\Omega_{\rm mDER}$ is most strongly constrained. 
At the other end, Quintessence has the sharpest slope for its dynamical component $\Omega_{\rm scf,\, ke}$ and its contribution rapidly diminished with redshift, leading to the loosest constraints. 
Current (and indeed future) data however will not be able to distinguish these models via $E(z)$ and $w(z)$ as shown in Fig.~\ref{fig:E_w_of_z_now}. 
Not only do DERs and Quintessence overlap in $E(z)$, but marginalising over \LCDM\ parameters further washes out any features at the background level.

Quintessence and DERs  
also
cluster differently. 
This leads to a slightly smaller matter clustering amplitude $\sigma_8$ at the $\leq 0.5 \sigma$ level in DER cosmologies, as well as in Quintessence and $w_0w_a$, which in turn results in slightly smaller $S_8 \equiv \sigma_8 \sqrt{\Omega_m/0.3}$ as shown in Table~\ref{tab:results_current}. 
Dark energy radiation can hence aid in mildly reducing the weak-lensing tension in cosmology wherein weak-leaning observations prefer smaller $S_8$ than predicted by the CMB using \LCDM\ \cite{DiValentino:2020vvd, Abdalla:2022yfr}. 
However, from forecasts we find that future CMB and SNe data will still not be able to distinguish between the two scenarios, in agreement with the conclusions of Ref. \cite{Wolf:2023uno} about the lack of discriminatory power of quintessence models in data.  
Any signals that may show up in the matter power spectra are currently well within error bars as shown in Fig.~\ref{fig:P_of_k_now}, particularly when marginalising over \LCDM\ parameters. 
Future data follows a similar pattern with marginally smaller error bands on $P(k)$. 
Note that these error bands are based on power spectra predicted from CMB, SNe and BAO data, 
and not a measurement of the power spectrum in the late universe. 
Predicting $P(k)$ requires the assumption of a cosmological model, and hence these error bands are model-dependent. 
They are tightest for \LCDM\ which has the fewest cosmological parameters as described above. 
These error bars are further dependent on the degeneracy directions in theory parameter space constrained by the combination of CMB+BAO+SNe data. 
Weak lensing data sets will have different degeneracy directions \cite{Secco:2022kqg} and future data from the Legacy Survey of Space and Time (LSST) at the Vera C. Rubin Observatory may provide additional insight into DER models, and constraints on DER may be a viable goal for future surveys. 

\section{Detection Prospects of Dark Energy Radiation Particle Content}
\label{sec:particle_stuff}

With current and future CMB and supernova data unable to distinguish between the various DER scenarios and Quintessence, we next turn to 
the direct-detection prospects of DER if one of the thermalized particles has BSM couplings to the Standard Model.
In particular, we consider three interesting scenarios, where either neutrinos, axions or dark photons populate DER with temperature $T_{\text{DER}}$.
Since we need a microphysical model to relate the DER density to a temperature, and to understand what particles are likely candidates to  become thermalized in DER, we use minimal DER, as described in Sec.~\ref{subsec:mDER}, as our concrete benchmark model.

Having performed a state-of-the-art analysis comparing the DER observables with current cosmological datasets in Sec.~\ref{sec:cosmology}, we find the $2\sigma$ upper bounds on the fractional energy density of minimal DER to be $\Omega^{\text{max}}_{\text{mDER}} = 0.03$ from Fig.~\ref{fig:tri_now_fut_de_params}\footnote{
See Table~\ref{tab:results_current} for the full posteriors in the appendices. 
}. 
We then solve for the DER temperature as
\begin{equation}
\label{eq:temp}
    T_{\text{DER}} = \left(\frac{30 \Omega_{\text{mDER}} \rho_c }{g_* {\pi}^2} \right)^\frac{1}{4} \, ,
\end{equation}
where $\rho_c = 3 M^2_{Pl} H^2_0 \approx 41 \, \text{meV}^4$ is the critical energy density of the Universe today. The energy distribution of DER is then determined by a blackbody as 
\begin{equation}
\label{eq:blackbody}
    \frac{d\Omega_{\text{mDER}}}{d\omega}\left(\omega \right) = 
    \frac{g_*}{2 \pi^2 \rho_c} 
    \frac{\omega^3}{e^{\frac{\omega}{T_{\text{DER}}}}-1}\, ,
\end{equation}
where $\omega$ is the energy of the relativistic particles populating DER, and the blackbody temperature $T_{\text{DER}}$. 
In each of the following subsections, we will consider a simple scenario for which we explicitly specify $g_*$, and the resulting maximum temperature $T_{\text{DER}}$, that could be relevant for direct detection. 

\subsection{Neutrinos }
\label{subsec:neutrinos}
One interesting possibility \cite{Berghaus:2020ekh} is that SM neutrinos become thermalized as a part of DER.
For both possible neutrino mass hierarchies the lightest mass eigenstate can be lighter than the cosmic neutrino background (C$\nu$B)
 temperature ($T_{\nu} \approx 0.16 \, \, \text{meV}$), allowing for one of the three species to be relativistic today.
If that species thermalized with the dark radiation, its temperature would increase to that of the DER temperature, $T_{\text{DER}}$. 
Within minimal DER, the thermalization could occur if a light right-handed neutrino is part of the dark radiation.

As highlighted in Sec. \ref{subsec:mDER}, the dark radiation can contain fermions.
If those fermions $\Psi^a$ are in the adjoint representation of the non-Abelian group, they could thermalize a right-handed neutrino $\nu_R$, which can thermalize the left handed SM neutrino through a mass mixing term
\begin{figure}[h]
    \centerline{\includegraphics[scale=0.55]{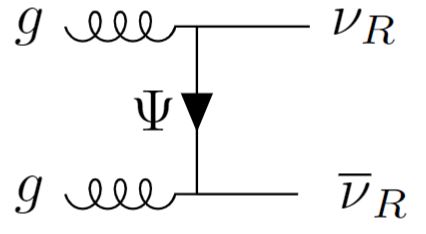}}
    \caption{ Production process of sterile neutrinos $\nu_R$ from a minimal dark radiation bath consisting of gauge bosons $g$ with a light fermion $\Psi$ in the adjoint representation. }
    \label{fig:Feynman}
\end{figure} 
\begin{equation}
    \mathcal{L}  \subset 
    - \frac{1}{f_{\nu_R}} G^a_{\mu \nu} \Psi^a \sigma^{\mu \nu} \nu_R  
    - y h \overline{\nu}_L \nu_R 
    - \frac{1}{2} m \overline\nu_R (\nu_R)^c 
    + h.c.  \, .
\end{equation}
Here we only consider mixing with the lightest SM neutrino, $\nu_L$, which we assume to have a Dirac mass term.
The size of the mass-mixing term is determined by the Yukawa coupling $y$, where $h$ is the SM Higgs field which obtains a vacuum expectation value $v$, $\langle h \rangle = v $.
In order for the right-handed neutrino to become thermalized, its mass $m$ has to be smaller than the temperature of the dark radiation bath ($m < T_{\text{DER}}$). 
Furthermore, the process shown in Fig.~\ref{fig:Feynman} of gauge bosons scattering into right-handed neutrinos, whose cross-section scales as $\sigma_{gg \to \nu_R \overline{\nu}_R}  \propto \frac{1}{f^2_{\nu_R}}$ needs to be efficient. 
Estimating the interaction rate as 
$\Gamma_{\text{int}} = \sigma n v $, with $n = \frac{\zeta(3)}{\pi^2} \left( \frac{g_{*}}{16} \right) \left( \frac{T_{\text{DER}}^3}{\text{meV}^3}\right)$, 
we find that the the interaction rate exceeds the expansion rate $H_0 \approx 1.5 \times 10^{-30} \,  \text{meV}$ of the Universe today, when $f_{\nu_R} \ll \text{TeV}$. 
The process for producing SM neutrinos is the same one as shown in Fig.~\ref{fig:Feynman} with an insertion of a mass-mixing term converting a sterile right-handed neutrino $\nu_R$ to a SM one $\nu_L$. 
Cosmological production of the right-handed neutrino species is suppressed in the early universe due to high temperatures such that $N_{\text{eff}}$ constraints are trivially avoided. 
Detailed constraints on the existence of light right-handed neutrinos with small mass splitting have been considered in \cite{Bakhti:2013ora,Chen:2022zts}, and can at most exclude mixing angles larger than $\theta = \frac{y h}{m} \lesssim 10^{-2}$. 
The cross-section for that production process in the late universe during dark energy domination scales as $\sigma_{gg \to \nu_R \nu_L}  \propto \frac{1}{f^2_{\nu_R}} \frac{y^2 h^2}{T^2_{\text{DER}}} $.
Then as long as 
\begin{equation}
    f_{\nu_R} \ll \text{GeV} \sqrt{{\frac{0.63 \, \text{meV}}{T_{\text{DER}}}}} \left( \frac{10^{-2} \,  \text{meV}}{y h} \right) \,   ,
\end{equation}
the SM neutrino will become thermalized as well. 
Thus, the thermalization of the sterile neutrino trivially allows for the possibility of a thermalized SM neutrino species. 
For this setup, we minimally require seven bosonic degrees of freedom from three gauge bosons with two polarizations for $N_c = 2$, plus one scalar degree of freedom for the axion, as well as fourteen fermionic degrees of freedom corresponding to two polarizations each for the left-handed SM neutrino, the right-handed neutrino, and the three fermions $\Psi^a$ in the adjoint representation of SU(2), which adds up to $g_* = 19.25$, and a maximum temperature of $T_{\nu \text{DER}} = 0.66 \, \text{meV}$. 
The amount of energy density in the thermalized left-handed neutrino (and anti-neutrino) is limited to $\Omega_{\nu \text{DER}} \leq  \frac{1.75}{g_*} \times \Omega^{\text{max}}_{\text{mDER}} = 0.0027$. 
We refer to this scenario as $\nu$DER.

\subsubsection{Prospects of Detection with Ptolemy}

The Ptolemy project \cite{PTOLEMY:2018jst, PTOLEMY:2019hkd} aims to directly detect the cosmic neutrino background (C$\nu$B) through capture on a tritium nucleus. 
The capture process
\begin{equation}
    \nu_e +{}^3\text{H} \to {}^3\text{He} +e^-    
\end{equation}
has a smoking-gun signature of a peak in the electron spectrum above the $\beta$-decay endpoint. 
The cosmic relic neutrino raises the center-of-mass (COM) energy of the process compared to the $\beta$ decay, allowing for ejected electrons with kinetic energies that exceed the kinetic energies possible in regular $\beta$ decays. 
In a standard cosmology, Ptolemy is most sensitive to massive non-relativistic neutrino species. 
The reason for this is two-fold. 
Massive neutrinos may cluster \cite{Zhang:2017ljh}, enhancing the average number density per neutrino state
\begin{equation} 
    \label{eq:relic CNB}
    n_0^{\text{C}\nu\text{B}} = \frac{3 \zeta(3)}{4 \pi^2} T^3_{\nu,0} = 56 \, \text{cm}^{-3}
\end{equation}
with $T_{\nu,0} = 1.95 \, \text{K} = 0.163 \, \text{meV}$ by up to $200\% \, (10\% - 20\%)$ for masses of $150 \, \text{meV}\, (60 \, \text{meV})$. 
This may allow for an increased capture rate compared to a relativistic neutrino with average number density $n_0$. 
The second reason pertains to the increase in COM-energy of the capture process compared to the $\beta$ decay. 
Non-relativistic relic neutrinos dominantly raise the COM-energy through the contribution of their rest mass, rather than their kinetic energy. 
Then the mass scale dictates where the peak above the $\beta$-decay endpoint lies, and crucially, whether it is resolvable with current technology. 

\begin{figure*}
    \centering    
    \includegraphics[width=0.49\textwidth]{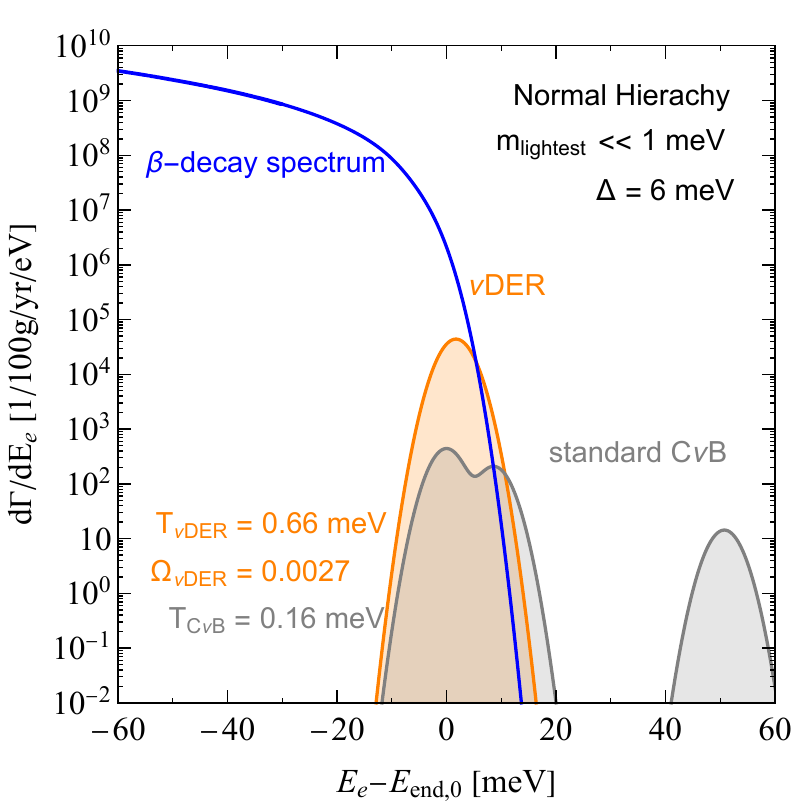}
    \includegraphics[width=0.488\textwidth]{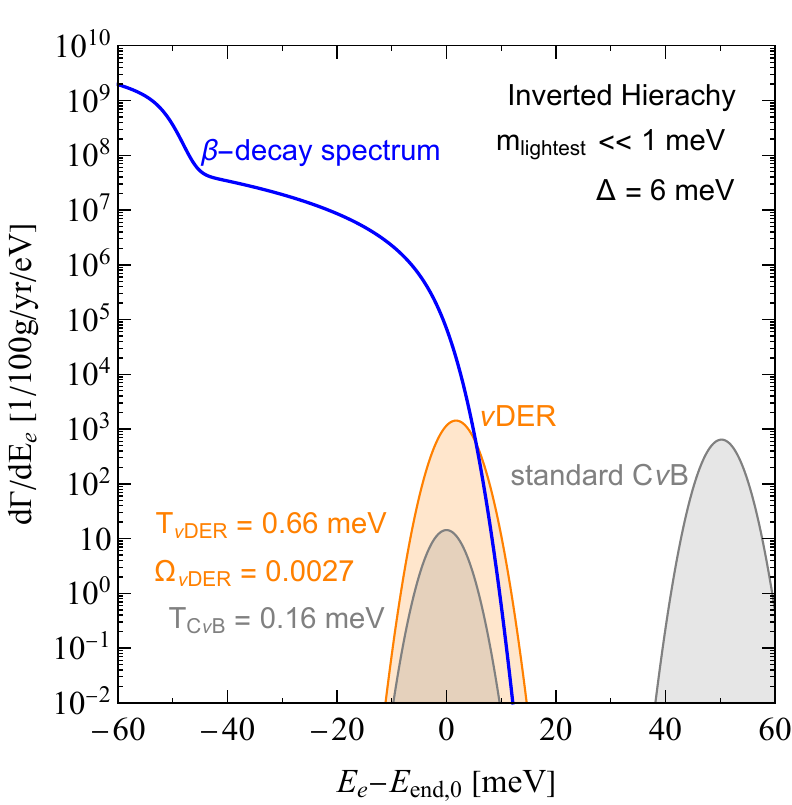}
    \caption{
    Expected event rates $d\Gamma/dE_e$ versus electron energy $E_e$ in a direct-detection experiment like Ptolemy (assuming 100g of tritium source) near the $\beta$-decay endpoint with an experimental resolution of $\Delta = 6 \, \text{meV}$ for a normal mass hierarchy (\textit{left}) and an inverted mass hierarchy (\textit{right}). The signal of the SM neutrino thermalized with DER ($\nu$DER) is shown in orange. The standard cosmic neutrino background (C$\nu$B) is contrasted in 
    gray, and the $\beta$-decay background spectrum is shown in blue.   }
    \label{fig:neutrinos}
\end{figure*}

In the $\nu$DER cosmology discussed above, with a relativistic SM neutrino that has thermalized with dark radiation to a temperature $T_{\nu\text{DER}}$, the prospects for the detection of a relativistic neutrino with Ptolemy change. 
The average number density per state increases by a factor of 80 as 
\begin{equation}
    n_0^{\nu\text{DER}} = \frac{3 \zeta(3)}{4 \pi^2} T^3_{\nu \text{DER}} = 4.4 \times 10^3 \,  \left(\frac{ T_{\nu \text{DER}}}{0.70 \, \text{meV}} \right)^3 \, \text{cm}^{-3}  \, , 
\end{equation}
Furthermore the scale that dictates the peak above the $\beta$-decay endpoint is given by $T_{\nu \text{DER}}$ rather than the negligible neutrino mass, and with a wider spread. 
The differential event rate for non-relativistic neutrino capture is determined by 
\begin{equation}
    \frac{d\Gamma}{dE_e} = \sum_i N_T |U_{ei}|^2 \overline{\sigma} n_0 \delta\left(E_e-(E_{\text{end,0}}+m_i)^2\right) \, ,
\end{equation}
where $N_T$ represents the number of tritium nuclei in a sample of mass $M_T$, $U_{ei}$ are the mixing-matrix elements, $m_i$ is the neutrino mass, and $E_{\text{end},0} = 19.0464 \, \text{keV}$ is the energy at the $\beta$-decay endpoint for massless neutrinos, and the average neutrino-capture cross-section is
\begin{equation}
    \overline{\sigma} = 3.8 \times 10^{-45} \text{cm}^{-2} \, , 
\end{equation}
approximately independent of the neutrino energy $E_\nu$.
The expected number of integrated signal events for the standard relic neutrino background for a 100g tritium detector per year is then around 4, assuming no enhancement through clustering. 
The average neutrino-capture cross-section for relativistic neutrinos is a factor of 2 larger than the non-relativistic one \cite{Long:2014zva}, which together with the factor of 80 enhancement in the number density for DER neutrinos would lead us to expect about 640 events.
We calculate the differential signal rate for the relativistic $\nu$DER neutrinos as
\begin{align}
    \frac{d\Gamma}{dE_e} = 2  N_T |U_{e,{\nu}}|^2  \overline{\sigma} & \int
    \frac{d^3p_{\nu}}{(2 \pi)^3} \frac{p_{\nu}^2}{e^{E_{\nu}/T_{\text{DER}}}+1} \\ 
    &\times \delta\left(E_e-(E_{\text{end,0}}+E_{\nu})^2\right) \, , 
\end{align}
and convolve the resulting spectrum with a Gaussian of full width at half maximum given by $\Delta$ to account for experimental resolution as described in detail in \cite{PTOLEMY:2019hkd}.
In Fig.~\ref{fig:neutrinos}, we compare the convolved spectrum of the thermalized DER neutrinos ($\nu$DER) with a convolved spectrum of the standard neutrino background (C$\nu$B) with a relativistic neutrino much lighter than $1 \, \text{meV}$ for an experimental resolution of $\Delta = 6\, \text{meV}$. 
We consider the normal hierarchy (NO, left) in which the neutrino masses are $m_1 \approx 0 \, \text{meV}$, $m_2 = 8.6 \, \text{meV}$, $m_3 = 51 \, \text{meV}$, and the inverted hierarchy (IH, right) where the neutrino masses are $m_1 = 50 \, \text{meV}$, $m_2 = 0.051 \, \text{meV}$, $m_3 \approx 0 \, \text{meV} $. 
Fig.~\ref{fig:neutrinos} illustrates the large boost of expected events for $\nu$DER compared to the standard $\text{C}\nu\text{B}$.
The normal mass hierarchy predicts a larger rate due to the mixing-matrix elements between the electron neutrino and the lightest neutrino being larger in this hierarchy ($|U_{e1}|^2 = 0.66$) than the inverted one ($|U_{e3}|^2 = 0.022$).
For the displayed resolution of $\Delta = 6\, \text{meV}$ in Fig.~\ref{fig:neutrinos}, a portion of the $\nu\text{DER}$ events are distinguishable from the $\beta$-decay background.
To evaluate the sensitivity on experimental resolution, we compute the number of $\nu\text{DER}$ signal events per 100g tritium per year detected in the region where the signal dominates over the $\beta$-decay spectrum (defined where the blue and orange curves intersect), and compare to the number of $\beta$-decay background events. 
The results are displayed in Tab.~\ref{tab:neutrinos}. 
In particular, the predictions for a normal neutrino mass hierarchy look somewhat promising with 12 signal events and 7 background events, for an achieved resolution of $6\, \text{meV}$. 
The most optimistic energy resolution discussed by the Ptolemy collaboration \cite{PTOLEMY:2019hkd} in 2019 was $10 \,\text{meV}$, though those projections preceded the realization that quantum uncertainties limit the resolution possible by mounting tritium onto graphene sheets due to additional smearing in the $\beta$-decay spectrum  of O($100 \, \text{meV}$) \cite{Cheipesh:2021fmg,PTOLEMY:2022ldz}.
A Ptolemy-like detector with an energy resolution even smaller than $10 \, \text{meV}$ would be ideal for testing $\nu \text{DER}$, and as shown in Fig.~\ref{fig:neutrinos}, even a factor of two improvement would lead to $\mathcal{O}$(10) events.
However, there are significant technological challenges to achieve this resolution that must be overcome. 
One promising idea may be spin-polarized tritium which has the potential to reduce the quantum uncertainty to $6 \, \text{meV}$ \cite{Cavoto:2022xwo,Tullynew}, meeting the conditions necessary to detect $\nu \text{DER}$. 
The feasibility of this undertaking is subject to further research and development. 

\begin{table}
\begin{tabular}{ |p{1.5cm}|p{1cm}| p{1cm}|p{1cm}|p{1cm} |p{1cm}| p{1cm}|  }
    \hline 
    & \multicolumn{3}{|c|}{NO} & \multicolumn{3}{|c|}{IO} \\
    \hline
    $\Delta$ [\text{meV}] &S  & B & S/B &S & B &S/B \\
    \hline
    2 & 187   & 22  & 7  & 6  & 0.7 & 9 \\
    4 & 71   & 27   & 2.7  & 2.3  & 0.9 & 2.7 \\
    6 & 12   & 7  & 1.7  & 0.40  & 0.23 & 1.7 \\
    8 & 1   & 0.7   & 1.4  & 0.03 & 0.02 & 1.4  \\
    10&  0.04   & 0.03   & 1.3  & $ 0.001$& $0.0007$ & 1.9 \\
    \hline
\end{tabular}
\caption{
    Signal and background events for a fictional 100g tritium detector with experimental resolution $\Delta$ for a normal (NO) and inverted (IO) neutrino mass hierarchy. 
    }
\label{tab:neutrinos}
\end{table}
             
\subsection{Axions  }
\label{subsec:axions}

Another intriguing possibility is the detection of a relativistic axion background. Axions automatically become thermalized in DER as discussed in Sec.\ref{subsec:mDER}, and are therefore a natural candidate for direct detection. 
In a minimal realization of minimal DER, the degrees of freedom in the radiation could be as small as $g_* = 7$, accounting for two polarizations of three gauge bosons and the scalar degree of freedom of the axion. Plugging $g_* = 7$ into eq.~\eqref{eq:temp} we find that the temperature could be as large as $T_{\varphi\text{DER}} =0.84 \, \text{meV}$, with a total energy density in relativistic axions being $\Omega_{\varphi \text{DER}} = \frac{1}{g_*} \times \Omega^{\text{max}}_{\text{mDER}} = 0.0043$. 

If the axion background couples to the visible sector, for example via an axion-photon coupling,  
\begin{equation}
    \label{eq:photoncoupling}
    \mathcal{L} \supset 
    - \frac{g_{\varphi \gamma \gamma }}{4} \varphi \tilde{F}_{\mu \nu}F^{\mu \nu} 
    = g_{\varphi \gamma \gamma } \varphi \mathbf{E} \cdot \mathbf{B} \, ,
\end{equation}
its presence could source electric or magnetic fields in dark matter direct-detection experiments. 
However, there are stringent constraints on the size of the photon-axion interaction $g_{\varphi \gamma \gamma}$,
\begin{equation}
    g_{\varphi \gamma \gamma} \lesssim g^{\text{SE}}_{\varphi \gamma \gamma} = 0.66 \times 10^{-10} \, \text{GeV}^{-1} \,,
\end{equation} 
from star-emission bounds that prohibit excess cooling of compact astrophysical objects through axion emission \cite{CAST:2017uph, Ayala:2014pea,Carenza:2020zil}.\footnote{We estimate that the back-reaction from an axion-photon coupling of this size on the evolution of dark energy is highly suppressed in the regime we consider, where $g_{\varphi gg} \gg g_{\varphi \gamma \gamma}$. Furthermore,  a tachyonic mode instability which could greatly enhance photon particle production is prevented by the photon's plasma mass $m_{\gamma} > 10^{-11} \, \text{meV}$.  }

\begin{figure}[h]                   
\centerline{\includegraphics[scale=0.7]{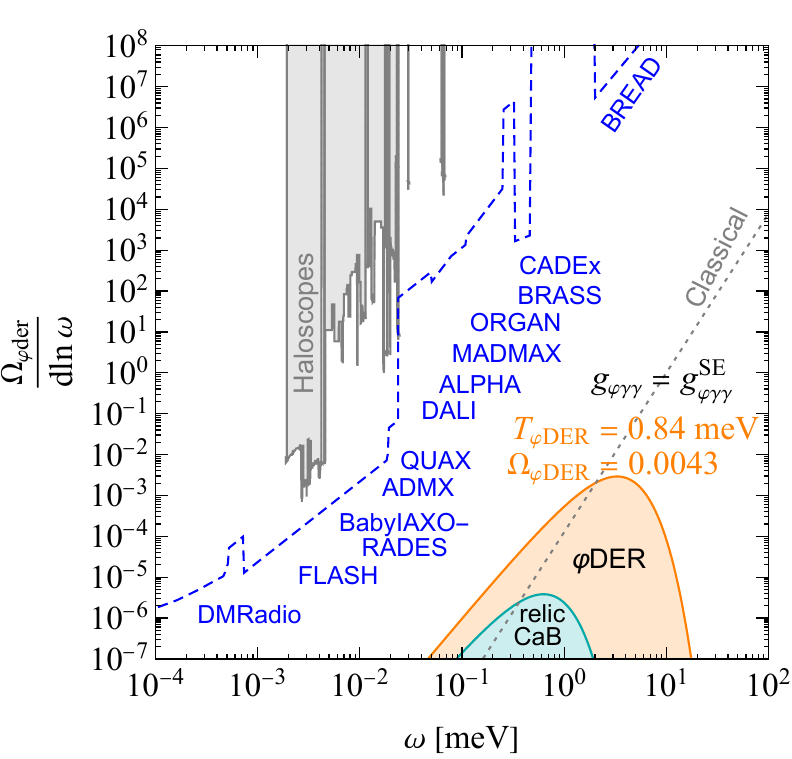} }
    \caption{ 
    An estimate of the sensitivity reach of various dark matter axion experiments \cite{AxionLimits} to relativistic axions with axion-photon coupling $g_{\varphi \gamma \gamma} = 0.66 \times 10^{-10} \, \text{GeV}^{-1}$. 
    The blue dashed lines indicate the sensitivity that could be reached with the next generation of experiments \cite{Caldwell:2016dcw, Lawson:2019brd, Alesini:2020vny, BREAD:2021tpx, PhysRevD.102.043003, PhysRevLett.123.141802, DMRadio:2022jfv, Aja:2022csb, Ahyoune:2023gfw, DeMiguel:2023nmz}. 
    The gray region indicates estimated constraints from existing data \cite{ADMX:2001nej, ADMX:2018gho, ADMX:2019uok, ADMX:2023rsk, HAYSTAC:2018rwy, HAYSTAC:2020kwv, HAYSTAC:2023cam, Salemi:2021gck}.  
    The orange region indicates the axion background predicted by minimal DER corresponding to a temperature$T_{\varphi \text{DER}} = 0.84 \, \text{meV}$ and an energy density in the axion background of $\Omega_{\varphi \text{DER}} = 0.0043$. 
    Shown in cyan is the allowed relic cosmic axion background (CaB), constrained to contribute less than $\Delta N_{\text{eff}} \leq 0.5$ to the radiation density in the early universe by the CMB, which  corresponds to $T_{\varphi} = 0.16 \, \text{meV}$. 
    The gray dotted line indicates the boundary at which it is reasonable to treat the axion background as a coherent classical wave (see text for details).
     }    
    \label{fig:axion}
\end{figure} 

In Fig.~\ref{fig:axion}, we show the estimated discovery potential of existing and proposed dark matter axion experiments to a relativistic thermal axion background when the axion-photon coupling saturates the star-emissions bound. 
Compared to the sensitivity to dark-matter-axion parameter space, there are several sources of suppression for a relativistic axion \cite{Dror:2021nyr}. 
The first one is the decrease in energy density of the relativistic species compared to the local dark matter density. 
The ratio of energy densities  is at best 
$\frac{\rho_{\varphi\text{DER}}}{\rho_{\text{DM}}} \simeq 10^{-7}$, 
with $\rho_\text{DM} \simeq 0.4 \, \text{GeV}/ \text{cm}^3$, and $\rho_{\varphi\text{DER}} = 0.0043 \times \rho_c$, where $\rho_c = 3 M^2_{\text{Pl}}H^2_0$ is the critical density. 
The second challenge to detectability lies in the broad energy distribution of the thermal axion $\frac{\Delta \omega_{\varphi\text{DER}}}{\omega_{\varphi\text{DER}}}\sim 1$, 
which  further limits the signal-power sensitivity of the detector, compared to axion dark matter with a long coherence time and a narrow energy distribution $\frac{\Delta \omega_{\text{DM}}}{\omega_{\text{DM}}} \sim 10^{-6}$. 
A dedicated study of relativistic axion direct-detection prospects, focused on DMRadio, ADMC and HAYSTAC  \cite{Dror:2021nyr}, found that this difference in energy distributions leads to a typical suppression factor due to the differences in bandwidth, 
$R_{\text{bw}}= \left(\frac{ \omega_{\text{DM}}}{\Delta \omega_{\text{DM}}} \frac{\Delta \omega_{\varphi}}{\omega_{\varphi}} \right)^{1/2}\sim 10^3$. 
For experimental designs targeting the larger DM axion masses ($m_a \gtrsim 0.1 \, \text{meV}$), the target DM wavelength is typically much smaller than the size of the detectors $\lambda_a \ll L$, which requires spatial power combining to acquire competitive sensitivity to a DM signal. This strategy is contingent on the spatial coherence of the DM signal, which is not given for our relativistic signal. Thus, the relativistic signal is subject to another suppression factor related to the collection efficiency which scales with the area of the sensor over the area of the detector, $R_{\text{ce}} = \frac{A_{\text{sensor}}}{A_{\text{detector}}} \sim 10^3$ \cite{Chounew}. We estimate the sensitivity to $\rho_{\varphi\text{DER}}$ shown in Fig.~\ref{fig:axion} as 
\begin{equation}
    \frac{\rho_{\varphi\text{DER}}}{\rho_{\text{DM}}} = R_{\text{bw}} R_{\text{ce}} \left(\frac{g_{\varphi \gamma \gamma}^{\text{lim}}}{g_{\varphi \gamma \gamma}}  \right)^2 \, , 
\end{equation}
taking the benchmark values \cite{Chounew} $R_{\text{bw}} = 10^3, R_{\text{ce}} = 1$ for DMRadio, FLASH, BabyIAXO-RADES, ADMX, QUAX, and CADEx, and $R_{\text{bw}} = 10^3, R_{\text{ce}} = 10^3$ for DALI, ALPHA, MADMAX and BRASS. 
For BREAD, which relies on threshold trigger detectors with non-frequency resolved readout, we use $R_{\text{bw}} = 1, R_{\text{ce}} = 10^7$.

Fig.~\ref{fig:axion} indicates the challenges of obtaining sensitivity to a thermal axion background. 
Despite the fact that DER allows for a thermal axion background with a  temperature five times larger than a relic axion background temperature, this enhancement is not large enough to overcome the relative suppression in energy density compared to the local dark matter density. 
Projections of future experimental sensitivities fall short of being able to directly detect $\varphi$DER in a dark-matter-axion experiment.      

Nevertheless, a thermal axion background is well motivated beyond the scope of DER, and justifies dedicated experimental efforts aiming to close the sensitivity gap for the detection of a relic CaB (indicated in Fig.~\ref{fig:axion} in cyan). 
Detection sensitivity to $\varphi$DER could be an interesting benchmark for such an initiative. 

\subsection{Dark Photons   }
\label{subsec:dphotons}

Simple extensions to minimal DER allow for dark photons to become thermalized with a temperature $T_{A'\text{DER}}$ \cite{Berghaus:2020ekh}. 
For example, if fermions $\psi$ charged under SU($N_c$) are part of the dark radiation, they could also be charged under a dark U(1) with fine-structure constant $\alpha'$.
Then scattering processes such as $g \psi \to A' \psi$ will also thermalize dark photons $A'$, as long as the production rate 
$\Gamma_{A'} \sim \alpha_D \alpha' T_{\text{DER}}$ 
is much larger than Hubble expansion, an easily full-filled requirement. 
If the dark photon $A'$ has a kinematic mixing $\kappa$ with the SM photon as in 
\begin{align}
    \label{eq:millicharge}
    \mathcal{L} \supset & 
        -\frac{1}{4}F_{\mu \nu} F^{\mu \nu} 
        +\frac{\kappa}{2} F_{\mu \nu} F^{'\mu \nu} 
        - \frac{1}{4}F_{\mu \nu}^{'}F^{'\mu \nu} \\ \nonumber
    & \frac{1}{2} m^2_{A'}A^{'}_{\mu}A^{'\mu} 
        + g'\bar{\psi} \gamma^{\mu} \psi A^{'}_{\mu} 
        + \bar{\psi}(i \partial - m_{\psi})\psi \,,
\end{align}
the fermions $\psi$ will have SM millicharge, and the dark photon may be detectable in direct-detection experiments. 
Of course, the fermions and dark photons can only be populated if they are light enough to constitute parts of DER, e.g. $m_{\psi}, m_{A'} \lesssim T_{A'\text{DER}}$.
Even without fermions initially present, dark photons may become populated through a higher dimensional operator such as
\begin{equation}
    \label{eq:FFdual}
    \mathcal{L} \supset 
    \frac{\alpha' \varphi}{16 \pi^2 f'}
    F^{'\mu \nu} \tilde{F}^{'}_{\mu \nu}   \,, 
\end{equation}
which allows for dark photon production through $gg \to A'A'$ with a rate that scales as $\Gamma_{A'} \sim \frac{T^5}{{(f'f)}^2}$, leading to thermalization if $f\sim f' \lesssim 10 \, \text{keV}$. 
Adding fermions only charged under the dark U(1) as explicitly written in Eq.~\ref{eq:millicharge} can make the thermalization more efficient \cite{Berghaus:2020ekh}, but is not necessary. 
There exists the possibility of a back-reaction to the evolution of the dark energy if the dark photon develops a tachyonic instability which happens when $\dot{\varphi}/f' > m_{A'}$.  
Here, we only consider the regime in which no significant particle production occurs from the zero mode such that any such back-reaction is irrelevant, and the dark photon is fully thermalized.

We take the approximate $2\sigma$ upper limit on minimal DER $\Omega_{\text{DER}} = 0.03$, and assume the minimal scenario where the degrees of freedom contributing to the radiation bath are $g_* =10$ for $N_c = 2$ including three gauge bosons with two polarizations, the axion with one degree of freedom, and a dark photon with three polarizations $g_A'$.
We then find the maximum temperature to be $T_{A'\text{DER}} = 0.78 \, \text{meV}$, with $\Omega_{A'\text{DER}} = \frac{3}{g_*} \times \Omega^{\text{max}}_{\text{mDER}} = 0.009$.

\subsubsection{Dark Photon Direct Detection Prospects}

Directly detecting a thermal dark photon background is an exciting prospect, and there are many upcoming experiments that will probe the parameter space populated by minimal DER.  
In this subsection, we will first discuss the viability of the size of the kinetic-mixing parameter $\kappa$, and will then estimate the reach of current and upcoming experiments.

\begin{figure*}
    \centering
    \includegraphics[width=0.49\textwidth]{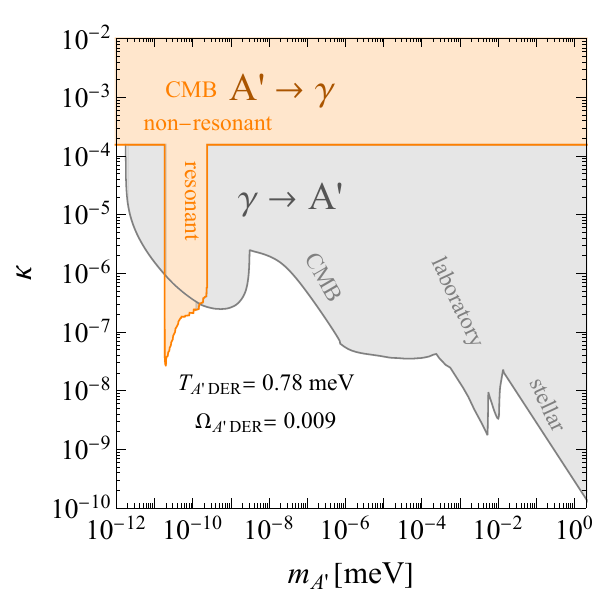}
    \includegraphics[width=0.49\textwidth]{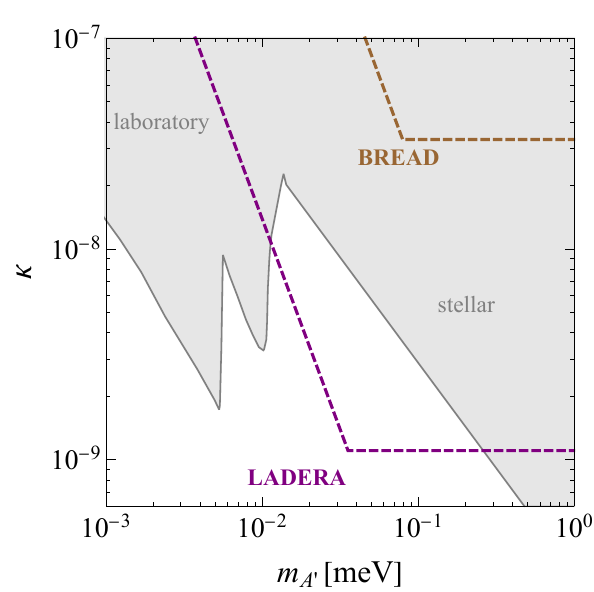}
    
    \caption{\textit{Left}  
    In gray and orange we show constraints on the dark photon-photon mixing parameter $\kappa$ as a function of the dark photon mass $m_{A'}$. 
    Orange shows bounds derived in this work from dark photon-to-photon conversion using FIRAS measurements of the CMB, 
    in the presence of a thermal dark-photon background with temperature $T_{A'\text{DER}} = 0.78 \, \text{meV}$, 
    the upper limit allowed by minimal DER. 
    In gray, we indicate the existing constraints on photon-to-dark photon conversions \cite{Caputo:2021eaa} 
    from the CMB \cite{Fixsen:1996nj, Mather:1998gm, Mirizzi:2009iz, McDermott:2019lch, Caputo:2020bdy, Caputo:2020rnx}, 
    laboratory probes such as fifth-force \cite{PhysRevLett.61.2285, Kroff:2020zhp} 
    and light-shining-through-a-wall experiments \cite{Ehret:2010mh, Romanenko:2023irv}, 
    as well as stellar-cooling arguments \cite{Schwarz:2015lqa, An:2020bxd, XENON:2021qze, Li:2023vpv}.  
    \textit{Right} 
    $1\sigma$ sensitivity forecast for detecting a dark photon background with temperature $T_{A'} =0.78 \, \text{meV}$ with LADERA \cite{Chounew} and BREAD \cite{BREAD:2021tpx}, assuming a 3-year exposure. Details in the text. }  
       \label{fig:photon}
\end{figure*}

Kinetic mixings with dark photons, regardless of the presence of a thermal background, are subject to various robust constraints \cite{Caputo:2021eaa} 
from laboratory probes such as fifth-force \cite{PhysRevLett.61.2285, Kroff:2020zhp} 
and light-shining-through-a-wall experiments \cite{Ehret:2010mh, Romanenko:2023irv}, 
as well as CMB measurements \cite{Fixsen:1996nj, Mather:1998gm, Mirizzi:2009iz, McDermott:2019lch, Caputo:2020bdy, Caputo:2020rnx}, 
and stellar-loss arguments \cite{Schwarz:2015lqa, An:2020bxd, XENON:2021qze, Li:2023vpv}, 
whose strength depend on the dark photon mass $m_A'$.
We indicate the bounds that only rely on photon-to-dark photon conversion, $\gamma \to A'$, in Fig. \ref{fig:photon} in gray. 

Additionally, the presence of a thermal background of kinetically-mixed dark photons can distort the CMB through non-resonant and resonant oscillations into SM photons. 
The Far Infrared Absolute Spectrophotometer (FIRAS) measured the shape of the CMB spectrum and found it fit extremely well to a blackbody spectrum at a temperature $T_{\gamma} = 2.725 \, \text{K}\, = 0.23 \, \text{meV}$ to an accuracy of $\simeq 10^{-4}\,  T_{\gamma}$ in the frequency range $0.284  \, \text{meV} < \omega_\gamma < 2.65  \, \text{meV}$ \cite{Fixsen:1996nj, Mather:1998gm}.

We estimate the non-resonant constraint by requiring
\begin{equation}
    \frac{\kappa^2 \frac{d\Omega_{A'}}{d\omega}(\omega,T_{A'\text{DER}})} 
    {\frac{d\Omega_\gamma}{d\omega}(\omega,T_{\gamma})} 
    \leq 10^{-4}
\end{equation}
in the measured range. 
Since the dark photon blackbody has a higher temperature than the CMB, the largest measured frequency provides the most stringent bound corresponding to $\kappa < 2 \times 10^{-4}$ for the benchmark $T_{A'\text{DER}} = 0.78 \, \text{meV}$, and $\Omega_{A'\text{DER}} = 0.009$. 
A PIXIE-like successor to the COBE/FIRAS measurements with a temperature resolution of $10 \,  \mu\text{K}$ \cite{A_Kogut_2011, Kogut_2020} could tighten the bounds on $\kappa$ by a factor of three.

To quantify limits from resonance conversions, we follow the treatments in \cite{Mirizzi:2009iz, Caputo:2020bdy}.
The resonant conversion requires the CMB photon's evolving plasma mass to match the mass of the dark photon, which makes resonant conversion efficient in only a narrow dark photon mass range.
Moreover, the temperature $T_{A'\text{DER}}(z)$ of the dark photon bath decreases with higher redshift, unlike the CMB temperature. 
Hence, only resonant conversions over recent redshifts ($z \lesssim 10$) can contribute an appreciable fraction of SM photons in the frequency range covered by FIRAS. 
We require 
\begin{equation}
    \frac{
    \frac{d\Omega_{A'}}{d\omega} (\omega,T_A'(z(t_{\text{res}}))} {\frac{d\Omega_\gamma}{d\omega} (\omega,T_{\gamma}(1+z(t_{\text{res}})))} 
    P_{A' \to \gamma}(t_{\text{res}}) 
    \leq 10^{-4} \,,
\end{equation}
where $t_{\text{res}}$ is the time at which the plasma mass matches the dark photon mass, and $P_{A' \to \gamma}(t_{\text{res}})$ is the probability that a dark photon oscillates into a SM photon given by 
\begin{equation}
    P_{A' \to \gamma}(t_{\text{res}}) 
    = \frac{\pi {m^2_{\gamma}(t_{\text{res}})} \kappa^2}{\omega} 
    \left| \frac{d \ln m^2_{\gamma}(t)}{dt} 
    \right|_{t = t_{\text{res}}}^{-1} \,.
\end{equation}
We additionally require that $\omega/(1+z(t_{\text{res}}))$ falls within the FIRAS frequency range.
The lowest frequency bin delivers the most stringent constraints on resonant conversion. 
The redshift evolution of $T_{A'\text{DER}}(z)$, can be obtained by solving the coupled system of Eqs.~\eqref{eq:EOM1} and \eqref{eq:EOM2}. 
An analytical approximation of $T_{A'\text{DER}}(z)$ is given in Eq.~26 in \cite{Berghaus:2020ekh}. 
We quantify the photon plasma mass $m^2_\gamma(t)$ using the procedure described in \cite{Mirizzi:2009iz, Caputo:2020bdy}, and neglect plasma inhomogeneities. Combined constraints from non-resonant and resonant oscillations of A'DER dark photons into SM photons
are shown in orange in Fig. \ref{fig:photon}. 
We leave a careful analysis of the impact of including inhomogeneities in the electron number density \cite{Caputo:2020bdy,Pirvu:2023lch} on the parameter space of the DER dark photons to future work.   

Fig. \ref{fig:photon} illustrates that the constraints on $\kappa$ weaken for very light dark photons $m_{A'} \lesssim 10^{-12} \, \text{meV}$. 
The thermal dark photon background produced in minimal DER only requires $m_{A'} \lesssim T_{A'\text{DER}}$, such that it can be thermalized in DER. 
In principle there is no lower limit on $m_A'$, however as we will see small dark photon masses are challenging to probe in direct detection experiments as the dark photon decouples in the small mass limit. 

Trying to detect an abundance of non-relativistic dark photons is an active area of on-going research with a plethora of new proposals in recent years \cite{Caldwell:2016dcw, BREAD:2021tpx, PhysRevD.102.043003, PhysRevLett.123.141802, Godfrey:2021tvs, Beaufort:2023qpd}. 
Most efforts are directed towards detecting a narrow mass distribution that comprises all or some of the observed dark matter. 
Detectors with sensitivity to non-relativistic dark photons also have some sensitivity to relativistic dark photons, however the reach is not competitive. Unlike in the relativistic axion scenario discussed in Sec.~\ref{subsec:axions}, the sensitivity depends on the relativistic dark photon mass. In order for an experiment to detect the dark photon background, it has to oscillate into visible photons within the characteristic length scale of the experiment $L$. That probability is given by 
\begin{equation}
\label{eq:oscill}
P_{A'\to A} = 4 \kappa^2 \sin^2 \left(\frac{m^2_{A'} L}{4\omega} \right)\,  ,
\end{equation}
such that an estimate to the sensitivity of a dark photon background scales with the mixing parameter $\kappa$, the dark photon mass $m^{\text{lim}}$ that the direct detection experiment is optimized for, as well as the dark photon mass $m_{A'}$ of the dark photon background,
\begin{equation}
    \label{eq:reach}
    \frac{\rho_{A'DER}}{\rho_{\text{DM}}} \propto \left(\frac{\kappa^{\text{lim}}}{\kappa} \right)^2
    \left( \frac{\sin(\frac{m^{\text{lim}} L}{4})}{ \sin(\frac{m^2_{A'}L_{\text{eff}}}{4 \omega} )} \right)^2 
    \, .
\end{equation}
Here $L_{\text{eff}}$ denotes the effective propagation length for relativistic dark photons in a direct detection experiment.
We note that dark matter experiments with sensitivity to $m^{\text{lim}} \ll \, \omega \sim \text{meV}$, can only probe a thermal background if $m_{A'} < m^{\text{lim}} \ll  \omega$, resulting in a large hit to the sensitivity.   
Thus, the most promising dark matter direct detection experiments are the ones probing $m^{\text{lim}} \sim \omega \sim \, \text{meV}$, such as BREAD, or alternatively an experiment designed to be optimized to a thermal relativistic signal. 

Such an effort is currently underway at the Jet Propulsion Lab (JPL) in Pasadena lead by Pierre Echternach and Aaron Chou.
The experiment, LAte Dark Energy RAdiation (LADERA) \cite{Chounew}, relies on similar detector technology as BREAD, however it has a large number of photon sensors, increasing the effective sensor area by a factor of $10^4$ to $A_s = 0.01 \, \text{m}^2$. Since a thermal relativistic signal, unlike a dark matter signal, cannot be effectively focused on a small sensor area, this provides a significant improvement in sensitivity over BREAD. LADERA consists of a shielded box with highly reflective walls and features $10^4$ pixels of quantum capacitance detectors (QCDs) \cite{2021JATIS...7a1003E}, which are single-mode, single THz ($4 \, \text{meV}$) photon sensors that are read out with a Quantum Instrumentation Control Kit (QICK) developed by Fermilab \cite{Stefanazzi:2021otz}.
LADERA's sensitivity is optimized in the range in which $m^2_{A'} \gg 4 \pi \omega/L_{\text{eff}}$ such that Eq.~\eqref{eq:oscill} simplifies to $P_{A' \to A} = 2\kappa^2$, which for $L_{\text{eff}} = 100 \, \text{cm}$ corresponds to $ m_{A'} > 3.5 *10^{-2} \, \text{meV}$, after which the sensitivity falls off steeply as $(m_{A'}/m^{\text{th}}_{A'})^4$. Taking an individual detector dark count rate of $1 \, \text{Hz}$ as the benchmark, and calculating the expected signal rate as 
\begin{equation}
\Gamma = n_{A'DER} v A_s \,,
\end{equation}
we forecast LADERA's  $1\sigma$ sensitivity to $\kappa = 1.1 \times 10^{-9}$ for a  three-year exposure.
 
\section{Conclusion  }
\label{sec:conclusion}

This decade may be pivotal in gaining insight into the nature of dark energy. 
Numerous experiments are coming online that probe precision late-universe cosmological observables, stress-testing models of dark energy. 
Simultaneously, Earth-based direct-detection programs are advancing rapidly. 
In this work, we examine the cosmological and direct-detection prospects of a novel dark energy model, dark energy radiation (DER), in which dark energy is comprised of a slowly-rolling scalar field that sources thermal dark radiation through dissipation. 
Remarkably, DER predicts an abundance of a thermal bath of particles which can be probed non-gravitationally. 

We investigate two variations of DER - minimal DER where the thermal friction leading to particle production emerges from a microphysical theory of an axion coupling to gauge fields; and toy DER, a toy model with a constant dissipation term. 
Current state-of-the-art precision cosmological data allow for an abundance of dark energy radiation up to $\Omega_{\text{DER}} = 0.03$ as seen from 2D posteriors, which exceeds the amount of dark radiation usually considered compatible with cosmological data by three orders of magnitude. We find that the constraints are dominated by the background behavior of DER. 
As DER is dynamically produced in the late universe, it evades the usual bounds on dark radiation which stem from the early universe. 
Forecasts with CMB and supernova data show the potential to improve these constraints on DER abundance by a factor of three. 
Moreover, such dark energy radiation can slightly decrease $S_8$ and improve the weak-lensing tension in cosmology. 
However, we find that neither current nor future cosmological data sets considered here will distinguish mDER and tDER from each other, or from Quintessence, making direct-detection of DER via interactions with the Standard Model the most promising avenue for discovery.

Using minimal DER as a benchmark model, we propose extensions that allow for neutrinos, axions or dark photons to become thermalized, and evaluate the prospects of directly detecting a thermal background comprised of these particles. 
We find that neutrinos thermalized in DER can enhance the average relic neutrino number density for a relativistic species by a factor of 80; 
however, an experiment relying on neutrino capture on tritium will require a resolution of $6 \, \text{meV}$ to distinguish the thermal neutrinos from the $\beta$-decay background spectrum, a technologically challenging feat. 
A normal neutrino hierarchy is favorable compared to an inverted hierarchy. 

The prospects of detecting a relativistic axion background are slim as the existing experimental program targets dark matter detection. 
Detecting DER would require an experiment specialized to a thermal axion background to probe interesting parameter space. 
Compared to a relic thermal axion background, the DER signal can be up to three orders of magnitude larger with a peak around $3 \, \text{meV}$ rather than $0.7 \, \text{meV}$, constituting an easier intermediate target for such an undertaking.

Dark photons are the most interesting direct-detection target. 
After including all constraints on the dark photon mixing parameter $\kappa$ in the presence of a thermal background, we forecast that viable parameter space can be probed with LADERA \cite{Chounew}. 
This promising result 
motivates further research and development of experimental search strategies targeting relativistic signals.

\section*{Acknowledgments}
We thank Chelsea Bartram, Aaron S. Chou, David Cyncynates, Christopher Dessert, Peter W. Graham, Junwu Huang, D'Arcy Kentworthy, Ken Van Tilburg, Zachary J. Weiner, and Lindley Winslow for helpful discussions. 
We thank Aaron S. Chou for providing LADERA sensitivity estimates, and for extensive feedback on how to improve the sensitivity estimates to a relativistic signal for experiments that target larger axion mass ranges.  
Simulations in this paper use High Performance Computing (HPC) resources supported by the University of Arizona TRIF, UITS, and RDI and maintained by the UA Research Technologies department. 
The authors would also like to thank the Stony Brook Research Computing and Cyberinfrastructure, and the Institute for Advanced Computational Science at Stony Brook University for access to the high-performance SeaWulf computing system, which was made possible by a $\$1.4$M National Science Foundation grant ($\# 1531492$). 
K.B. acknowledges the support of NSF Award PHY2210533, and thanks the U.S. Department of Energy, Office of Science, Office of High Energy Physics, under Award Number DE-SC0011632 and the Walter Burke Institute for Theoretical Physics. 
TK was supported by 
NASA ATP Grant 80NSSC18K0694, funds provided by the Center for Particle Cosmology at the University of Pennsylvania,
the Simons Foundation and the Kavli Institute for Cosmological Physics at the University of Chicago through an endowment from the Kavli Foundation. 
T.B. was supported through the INFN project “GRANT73/Tec-Nu”, and by the COSMOS network (www.cosmosnet.it) through the ASI (Italian Space Agency) Grants 2016-24-H.0 and 2016-24-H.1-201. 
This work is partially supported by ICSC – Centro Nazionale di Ricerca in High Performance Computing, Big Data and Quantum Computing, funded by European Union – NextGenerationEU.

\appendix

\section{Full marginalized posteriors }

This appendix section presents our comprehensive posteriors. 
These can also be accessed directly from our MCMC chains made available on \href{https://github.com/KBerghaus/class_der/tree/main/MCMC_chains}{GitHub}.

\begin{figure*}
    \centering
    \includegraphics[width=1.\textwidth]{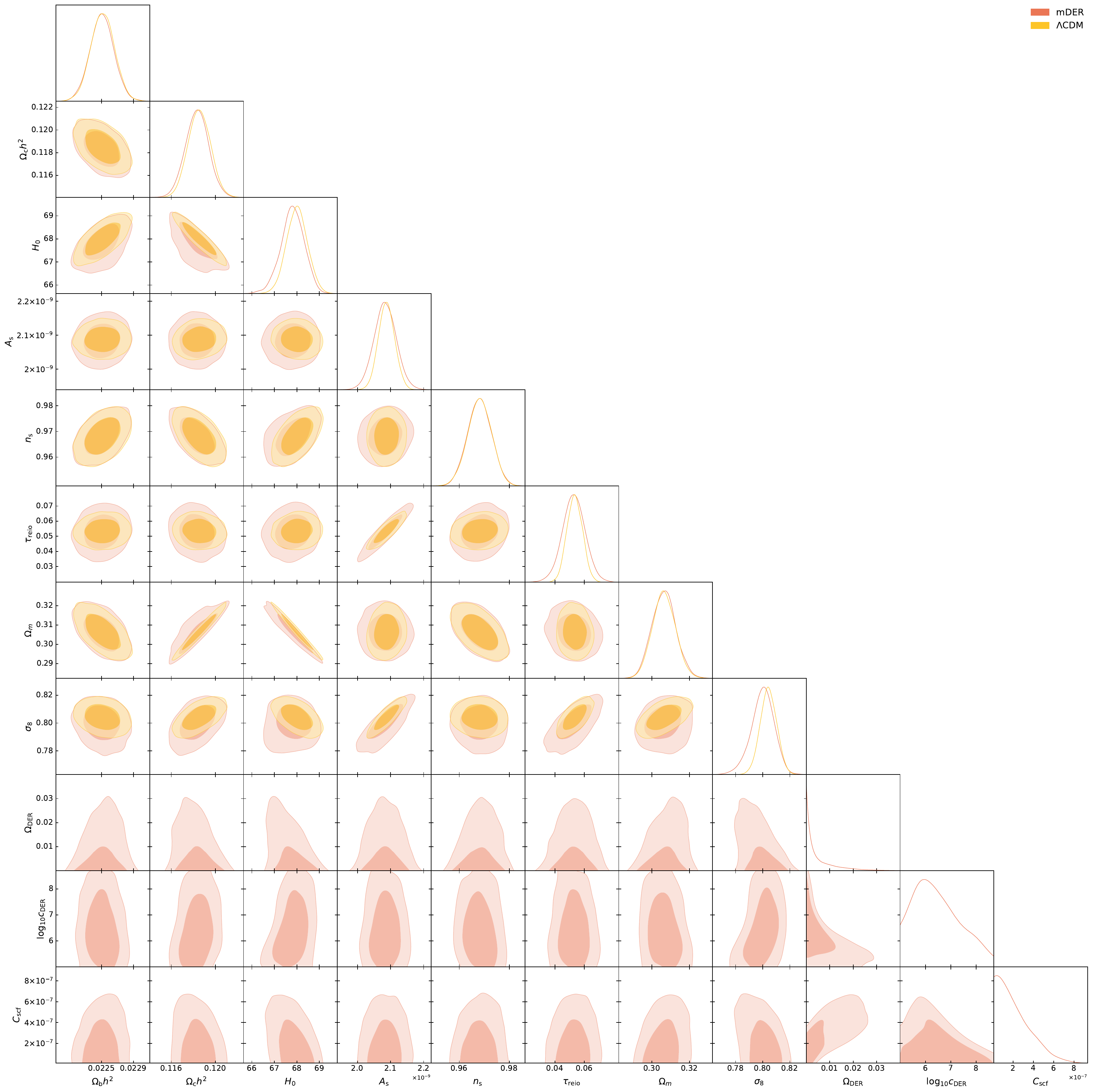}
    \caption{ 
    Constraints from current data for \LCDM\ (yellow) and minimal DER (orange). 
    We show the base \LCDM\ parameters along with a parameter $c_{\rm DER} = c_n$ controlling the thermal friction $\Upsilon = c_n \rho_{\rm DER}^{n/4}$, the slope $C_{\rm scf}$ of the scalar potential and the resultant contribution $\Omega_{\rm DER}$ of DER today. 
    For mDER, $n=3$. 
    }
    \label{fig:mDER_big_tri}
\end{figure*}

\begin{figure*}
    \centering
    \includegraphics[width=1.\textwidth]{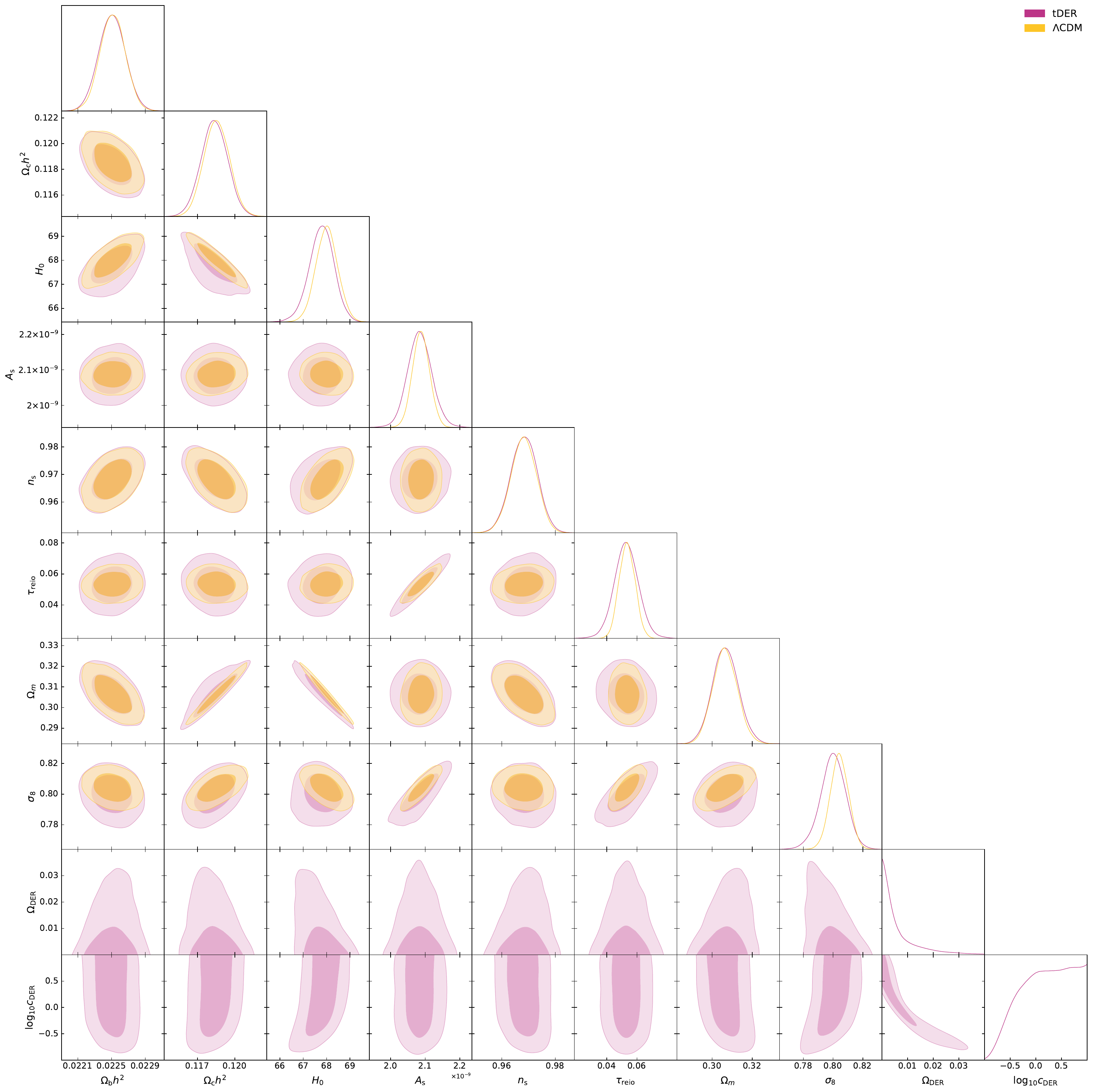}
    \caption{
    Constraints from current data for \LCDM\ (yellow) and toy DER (pink). 
    We show the base \LCDM\ parameters along with the thermal friction $\Upsilon = c_{\rm DER}$ and the resultant contribution $\Omega_{\rm DER}$ of tDER today. 
    }
    \label{fig:tDER_big_tri}
\end{figure*}

\begin{figure*}
    \centering
    \includegraphics[width=1.\textwidth]{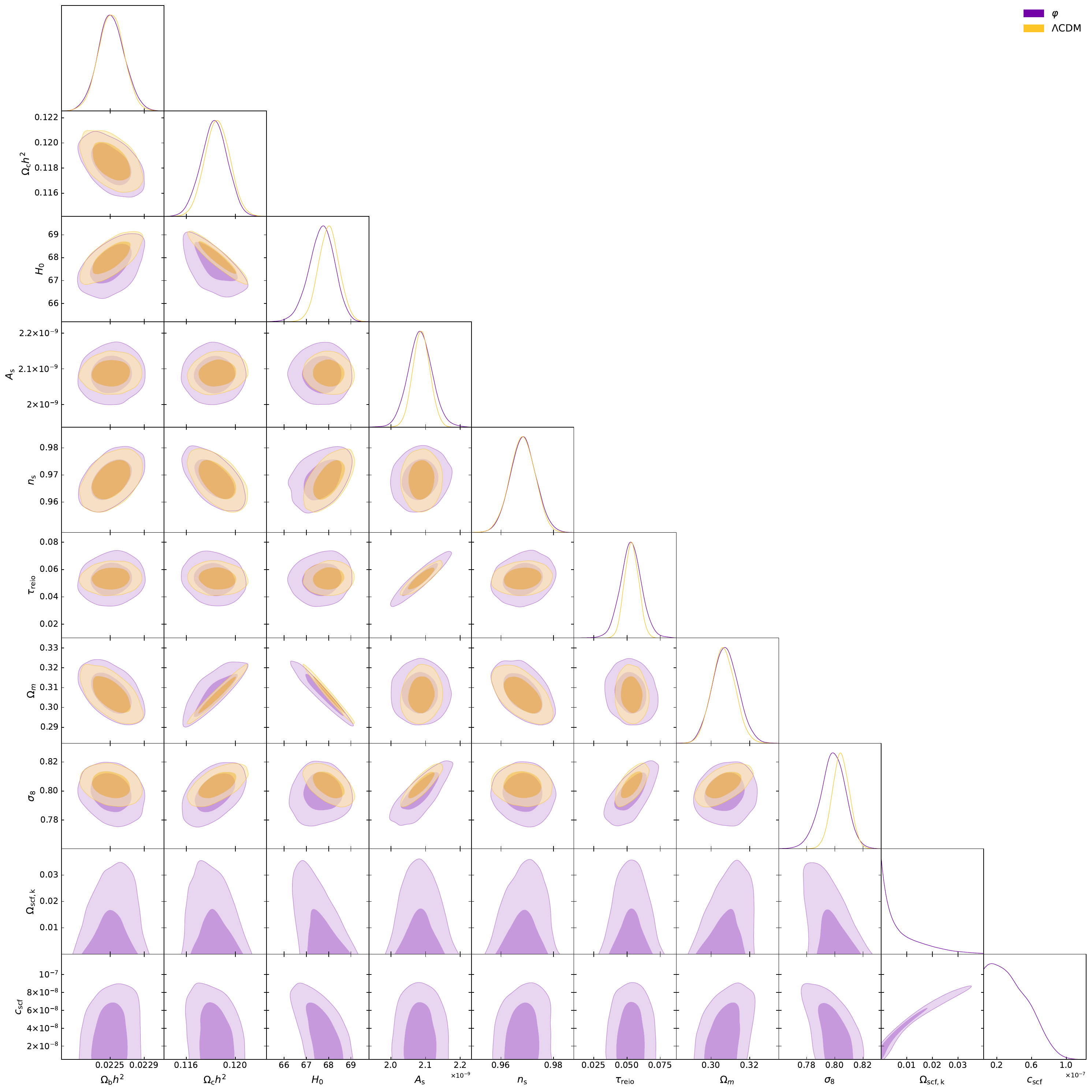}
    \caption{
    Constraints from current data for \LCDM\ (yellow) and Quintessence (purple). Besides the base \LCDM\ parameters, we also show the slope $C_{\rm scf}$ of the scalar potential and the contribution $\Omega_{\rm scf\,,k}$ of the kinetic energy of Quintessence today. 
    }
    \label{fig:quint_big_tri}
\end{figure*}

\begin{figure*}
    \centering
    \includegraphics[width=1.\textwidth]{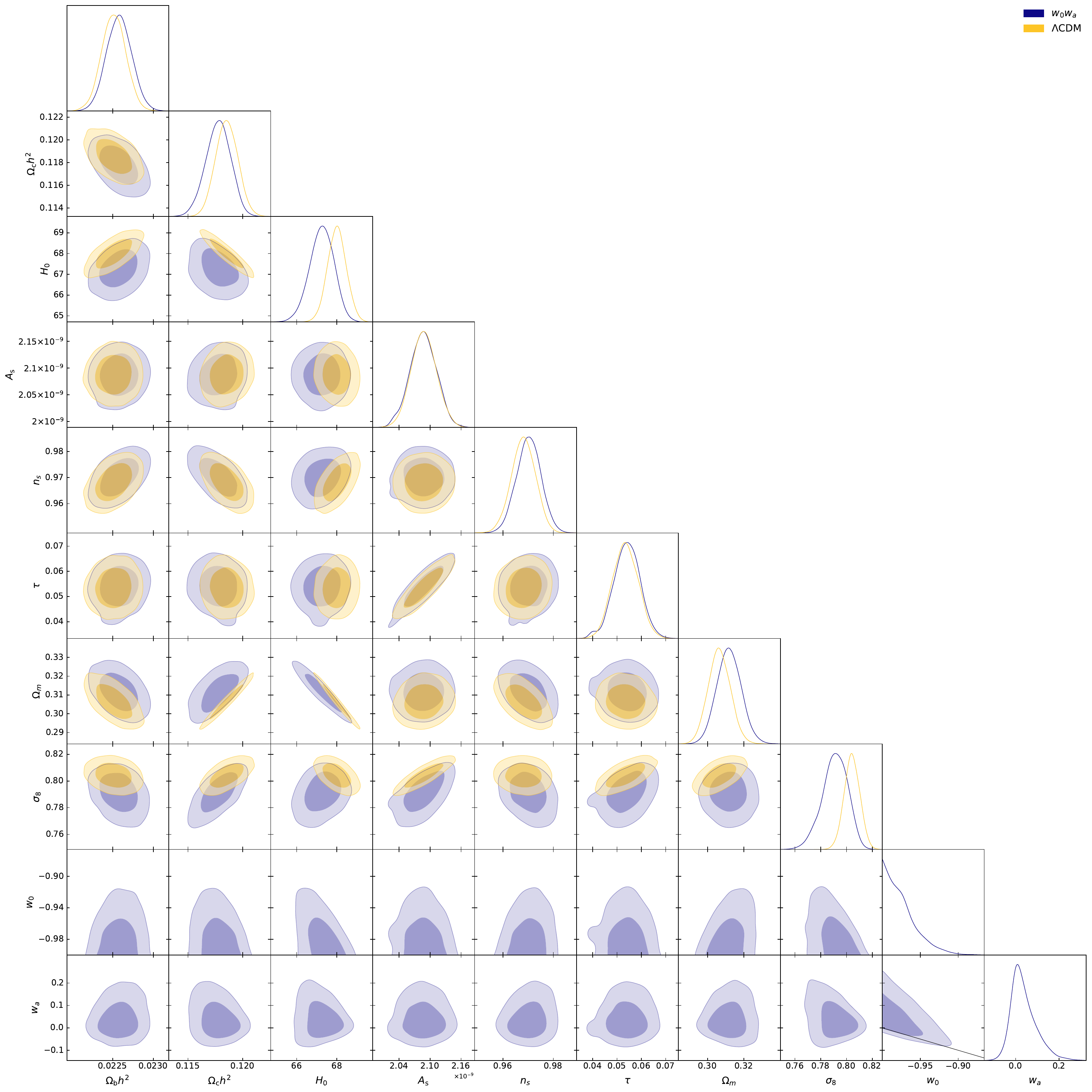}
    \caption{
    Constraints from current data for \LCDM\ (yellow) and $w_0w_a$ (blue). 
    Beyond the base \LCDM\ parameters, we show the equation of state of dark energy $w_0$ today and that in the deep past $w_a$. 
    The black solid line in the subplot showing $w_a$ vs. $w_0$ marks the boundary with the phantom region which is excluded by our priors. 
    The extent of the $2\sigma$ contour beyond this boundary is a feature of GetDist smoothing. 
    }
    \label{fig:w0wa_big_tri}
\end{figure*}

Table~\ref{tab:results_current} summarises the marginalised 1D posteriors for all cosmologies explored using our current data set, which includes BAO data from SDSS DR12 and DR7 main galaxy sample, Pantheon supernovae, and Planck 2018 CMB measurements of the low-$\ell$ TT and EE spectra and high-$\ell$ TTTEEE spectra as described in Sec.~\ref{subsec:data} upto $\ell < 1300$. 
Figs. \ref{fig:mDER_big_tri}-\ref{fig:w0wa_big_tri} show the 2D marginalised posteriors for cosmological parameters for mDER, tDER, Quintessence and $w_0w_a$ respectively. 

\begin{figure*}
    \centering
    \includegraphics[width=1.\textwidth]{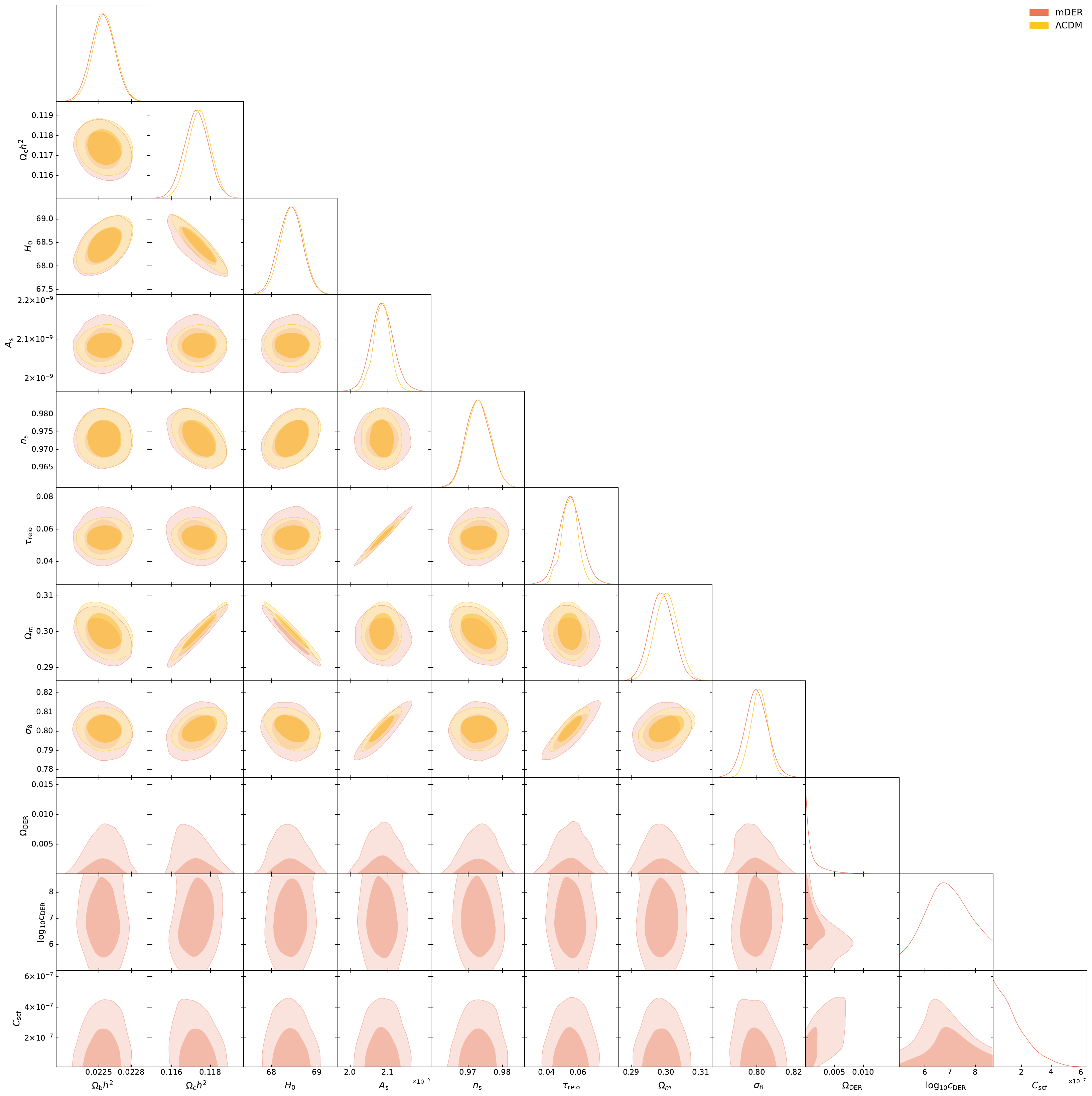}
    \caption{ 
    Constraints from future data for \LCDM\ (yellow) and minimal DER (orange), showing the base \LCDM\ parameters along with the parameter $c_{\rm DER} = c_n$ that controls the thermal friction $\Upsilon = c_n \rho_{\rm DER}^{n/4}$, the slope $C_{\rm scf}$ of the scalar potential and the resultant contribution $\Omega_{\rm DER}$ of DER today. 
    For mDER, $n=3$. 
    }
    \label{fig:mDER_big_tri_fut}
\end{figure*}

\begin{figure*}
    \centering
    \includegraphics[width=1.\textwidth]{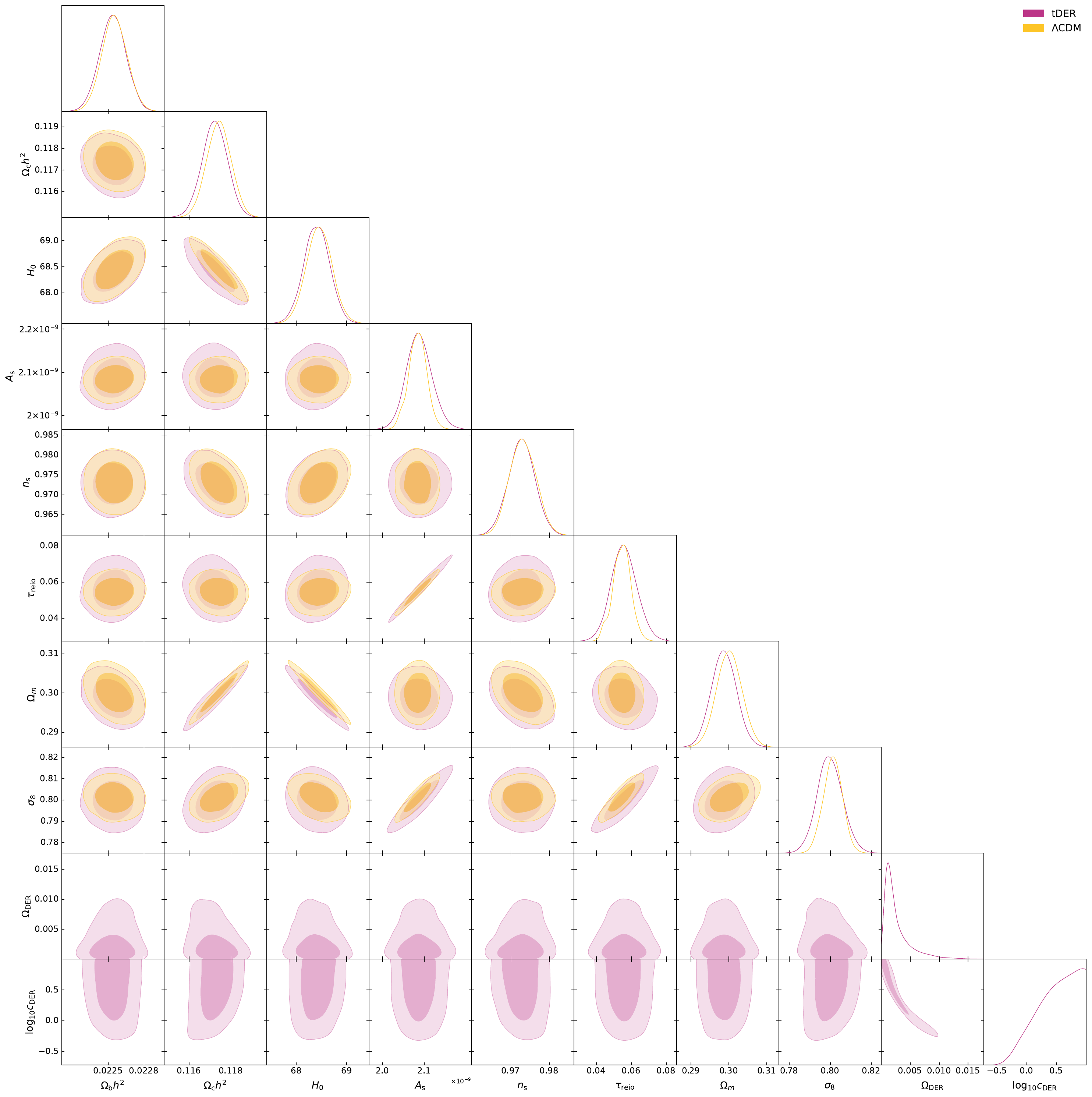}
    \caption{
    Constraints from future data for \LCDM\ (yellow) and toy DER (pink), showing the base \LCDM\ parameters along with the thermal friction $\Upsilon = c_{\rm DER}$ and the resultant contribution $\Omega_{\rm DER}$ of tDER today. 
    }
    \label{fig:tDER_big_tri_fut}
\end{figure*}

\begin{figure*}
    \centering
    \includegraphics[width=1.\textwidth]{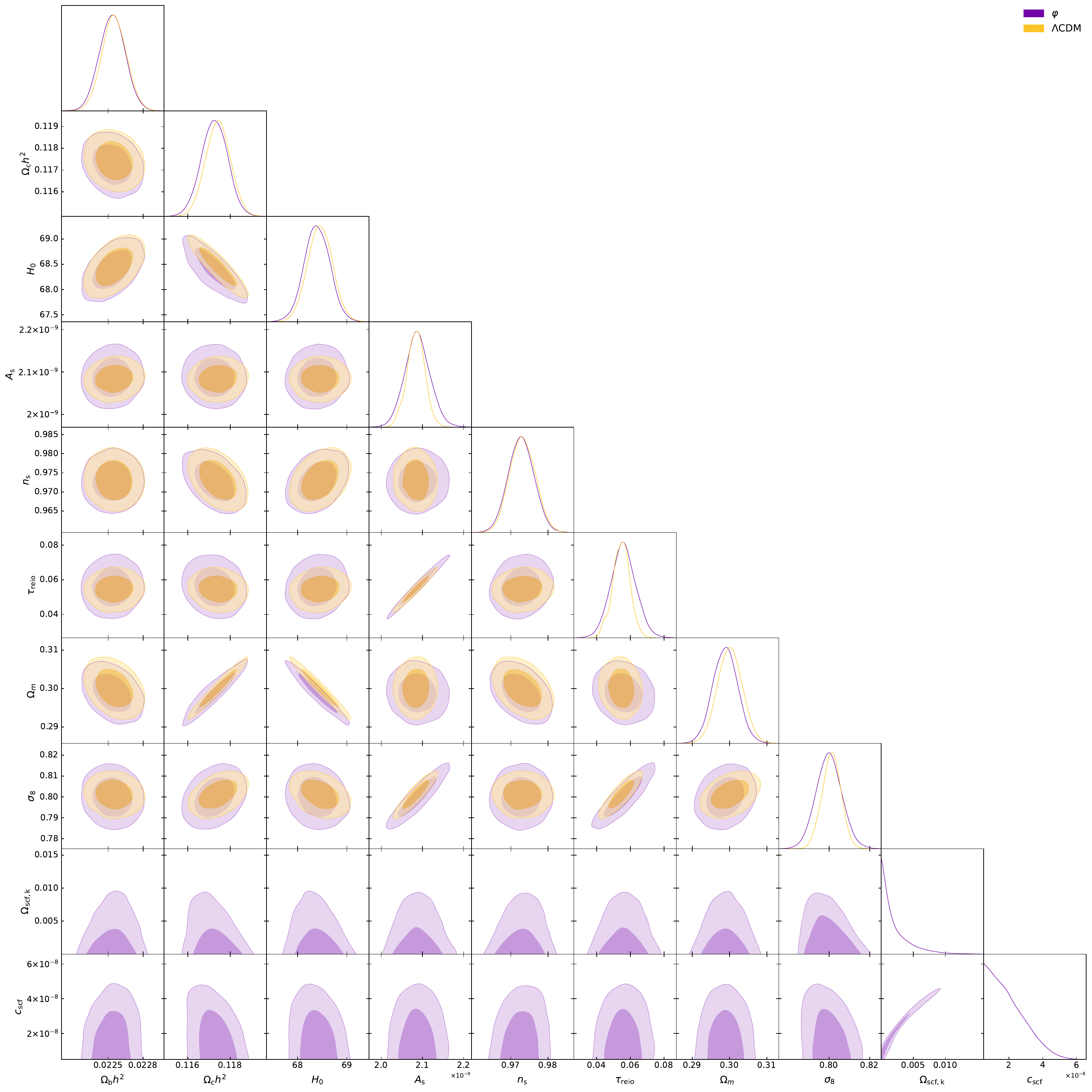}
    \caption{
    Constraints from future data for \LCDM\ (yellow) and Quintessence (purple). Besides the base \LCDM\ parameters, we show the slope $C_{\rm scf}$ of the scalar potential and the contribution $\Omega_{\rm scf\,,k}$ of the kinetic energy of Quintessence today. 
    }
    \label{fig:quint_big_tri_fut}
\end{figure*}

\begin{figure*}
    \centering
    \includegraphics[width=1.\textwidth]{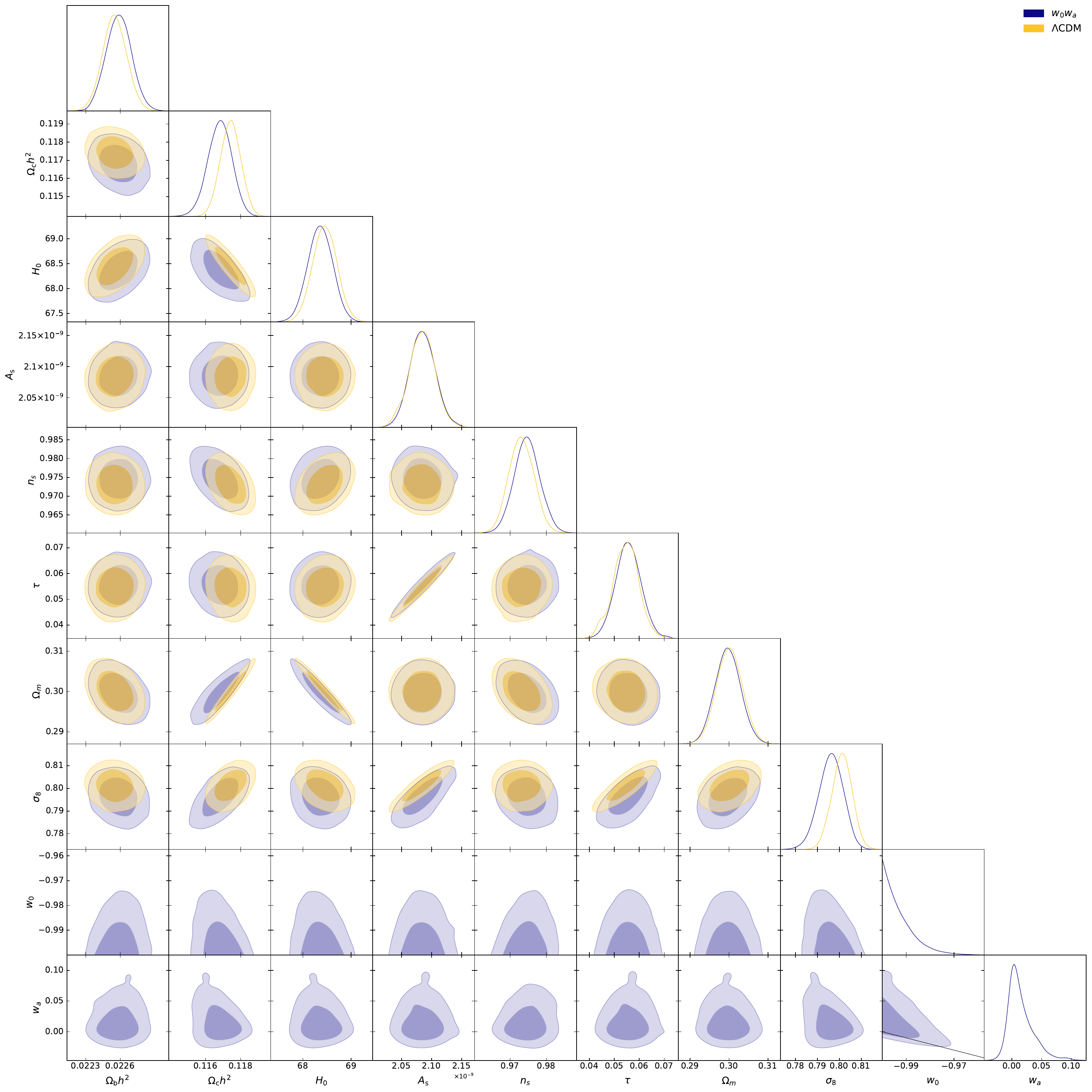}
    \caption{
    Constraints from future data for \LCDM\ (yellow) and $w_0w_a$ (blue). 
    Beyond the base \LCDM\ parameters, we show the equation of state of dark energy $w_0$ today and that in the deep past $w_a$. 
    The black solid line in the subplot showing $w_a$ vs. $w_0$ marks the boundary with the phantom region which is excluded by our priors. 
    The extent of the $2\sigma$ contour beyond this boundary is a feature of GetDist smoothing. 
    }
    \label{fig:w0wa_big_tri_fut}
\end{figure*}

Similarly, Table~\ref{tab:results_future} summarises the marginalised 1D posteriors for all cosmologies we explore using forecasts, which include BAO data from SDSS DR12 and DR7 main galaxy sample, Roman supernovae forecasts, Planck 2018 CMB measurements of the low-$\ell$ TT and EE spectra, and Simons Observatory projections for high-$\ell$ TTTEEE spectra upto $\ell < 1300$. 
Figs. \ref{fig:mDER_big_tri_fut}-\ref{fig:w0wa_big_tri_fut} show the corresponding 2D marginalised posteriors for cosmological parameters for mDER, tDER, Quintessence and $w_0w_a$ respectively.

\begin{table*}[]
    \centering
    \begin{tabular}{|l|c|c|c|c|c|}
\hline
Parameter   &   \LCDM   &   mDER    &   tDER    &   $\varphi$   &   $w_0w_a$ \\ 
\hline 
\hline
$\Omega_\mathrm{b} h^2$
	 & $ 0.02251\pm 0.00015$ 
	 & $ 0.02250\pm 0.00015$ 
	 & $ 0.02251\pm 0.00016$ 
	 & $ 0.02251\pm 0.00016$ 
	 & $ 0.02258\pm 0.00016$ 
	 \\
$\Omega_\mathrm{c} h^2$
	 & $ 0.1185\pm 0.0010$ 
	 & $ 0.1183\pm 0.0011$ 
	 & $ 0.1184\pm 0.0011$ 
	 & $ 0.1183\pm 0.0011$ 
	 & $ 0.1178\pm 0.0011$ 
	 \\
$H_0$
	 & $ 68.01\pm 0.47$ 
	 & $ 67.80^{+0.55}_{-0.48}$ 
	 & $ 67.78^{+0.55}_{-0.49}$ 
	 & $ 67.69^{+0.61}_{-0.53}$ 
	 & $ 67.25^{+0.64}_{-0.56}$ 
	 \\
$A_\mathrm{s} \times 10^{-9}$
	 & $2.088\pm 0.025$ 
	 & $2.084\pm 0.034$ 
	 & $2.084\pm 0.035$ 
	 & $2.085\pm 0.035$ 
	 & $2.087\pm 0.026$ 
	 \\
$n_\mathrm{s}$
	 & $ 0.9681\pm 0.0048$ 
	 & $ 0.9681\pm 0.0048$ 
	 & $ 0.9683\pm 0.0049$ 
	 & $ 0.9684\pm 0.0049$ 
	 & $ 0.9702\pm 0.0049$ 
	 \\
$\tau_\mathrm{reio}$
	 & $ 0.0535\pm 0.0052$ 
	 & $ 0.0528\pm 0.0078$ 
	 & $ 0.0529\pm 0.0081$ 
	 & $ 0.0530\pm 0.0081$ 
	 & $ 0.0540\pm 0.0055$ 
	 \\
$\Omega_m$
	 & $ 0.3065\pm 0.0061$ 
	 & $ 0.3064\pm 0.0063$ 
	 & $ 0.3067\pm 0.0065$ 
	 & $ 0.3074\pm 0.0068$ 
	 & $ 0.3118\pm 0.0068$ 
	 \\
$\sigma_8$
	 & $ 0.8043\pm 0.0061$ 
	 & $ 0.7999^{+0.0090}_{-0.0075}$ 
	 & $ 0.8000\pm 0.0085$ 
	 & $ 0.7987^{+0.0094}_{-0.0080}$ 
	 & $ 0.791^{+0.011}_{-0.0085}$ 
	 \\
$S_8$
	 & $ 0.813\pm 0.013$ 
	 & $ 0.808\pm 0.013$ 
	 & $ 0.809\pm 0.013$ 
	 & $ 0.808\pm 0.013$ 
	 & $ 0.806\pm 0.013$ 
	 \\
\hline \hline 
$\Omega_{\rm DER}$
	 & 	
	 & $< 0.0222$ 
	 & $< 0.0249$ 
	 & 	
	 & 	
	 \\
$\log_{10} c_{\rm DER}$
	 & 	
	 & $ 6.48^{+0.57}_{-1.2}$ 
	 & $> -0.595$ 
	 & 	
	 & 	
	 \\
$C_{\rm scf}$ $[M_{\text{pl}}\text{Mpc}^{-2}]$
	 & 	
	 & $< 5.42\times 10^{-7}$ 
	 & 	
	 & $< 7.57\times 10^{-8}$ 
	 & 	
	 \\
$\Omega_{\rm scf,k}$
	 & 	
	 & $< 3.04\times 10^{-4}$ 
	 & $< 6.35\times 10^{-5}$ 
	 & $< 0.0271$ 
	 & 	
	 \\
$w_0$
	 & 	
	 & 	
	 & 	
	 & 	
	 & $< -0.933$ 
	 \\
$w_a$
	 & 	
	 & 	
	 & 	
	 & 	
	 & $ 0.037^{+0.037}_{-0.067}$ 
	 \\

\hline \hline

$\chi^2$
	 & $ 3332.1\pm 5.5$ 
	 & $ 3332.0\pm 5.7$ 
	 & $ 3332.5\pm 5.7$ 
	 & $ 3332.7\pm 5.8$ 
	 & $ 3334.5\pm 5.8$ 
	 \\
$\chi^2_\mathrm{BAO\, DR12}$
	 & $ 4.16\pm 0.98$ 
	 & $ 4.08\pm 0.98$ 
	 & $ 4.1\pm 1.0$ 
	 & $ 4.03\pm 0.98$ 
	 & $ 4.3\pm 1.4$ 
	 \\
$\chi^2_\mathrm{BAO\ DR7\ MGS}$
	 & $ 1.57\pm 0.47$ 
	 & $ 1.48\pm 0.47$ 
	 & $ 1.47\pm 0.48$ 
	 & $ 1.45\pm 0.48$ 
	 & $ 1.14\pm 0.46$ 
	 \\
$\chi^2_\mathrm{SNe}$
	 & $ 1034.93\pm 0.22$ 
	 & $ 1035.25\pm 0.71$ 
	 & $ 1035.29\pm 0.75$ 
	 & $ 1035.6\pm 1.1$ 
	 & $ 1036.3\pm 1.3$ 
	 \\
$\chi^2_\mathrm{Planck\ lowTT}$
	 & $ 22.73\pm 0.93$ 
	 & $ 22.8\pm 1.0$ 
	 & $ 22.73\pm 0.95$ 
	 & $ 22.73\pm 0.93$ 
	 & $ 22.43\pm 0.89$ 
	 \\
$\chi^2_\mathrm{Planck\ lowEE}$
	 & $ 396.35\pm 0.88$ 
	 & $ 396.8\pm 1.5$ 
	 & $ 396.9\pm 1.7$ 
	 & $ 396.9\pm 1.7$ 
	 & $ 396.42\pm 0.93$ 
	 \\
$\chi^2_\mathrm{Planck\ high-\ell}$
	 & $ 1872.3\pm 5.4$ 
	 & $ 1871.5\pm 5.5$ 
	 & $ 1872.0\pm 5.5$ 
	 & $ 1872.0\pm 5.5$ 
	 & $ 1873.9\pm 5.8$ 
	 \\
\hline 
    \end{tabular}
    \caption{
    Marginalised 1D $1\sigma$ posteriors and means are shown for various cosmologies described in Section~\ref{subsec:models}, constrained using current data as described in Section~\ref{subsec:data}.  
    Here $c_{\rm DER} = c_3$ in units of $[ 3^{-\frac{3}{4}}M_{\text{pl}}^{-\frac{3}{2}} \text{Mpc}^{\frac{1}{2}}]$ for mDER and $c_0$ in units of $[\text{Mpc}^{-1}]$ for tDER. 
    For one-sided bounds, we show the 95\% upper or lower limit. 
    }
    \label{tab:results_current}
\end{table*}

\begin{table*}[]
    \centering
    \begin{tabular}{|l|c|c|c|c|c|}
\hline
Parameter   &   \LCDM   &   mDER    &   tDER    &   $\varphi$   &   $w_0w_a$ \\ 
\hline 
\hline
$\Omega_\mathrm{b} h^2$
	 & $ 0.02255\pm 0.00011$ 
	 & $ 0.02254\pm 0.00011$ 
	 & $ 0.02254\pm 0.00011$ 
	 & $ 0.02254\pm 0.00011$ 
	 & $ 0.02259\pm 0.00011$ 
	 \\
$\Omega_\mathrm{c} h^2$
	 & $ 0.11743\pm 0.00059$ 
	 & $ 0.11728\pm 0.00063$ 
	 & $ 0.11722\pm 0.00061$ 
	 & $ 0.11726\pm 0.00062$ 
	 & $ 0.11681\pm 0.00069$ 
	 \\
$H_0$
	 & $ 68.46\pm 0.25$ 
	 & $ 68.42\pm 0.26$ 
	 & $ 68.41\pm 0.25$ 
	 & $ 68.39\pm 0.26$ 
	 & $ 68.36\pm 0.26$ 
	 \\
$A_\mathrm{s} \times 10^{-9}$
	 & $2.084\pm 0.022$ 
	 & $2.085\pm 0.030$ 
	 & $2.088^{+0.028}_{-0.032}$ 
	 & $2.088\pm 0.031$ 
	 & $2.085\pm 0.022$ 
	 \\
$n_\mathrm{s}$
	 & $ 0.9731\pm 0.0034$ 
	 & $ 0.9729\pm 0.0035$ 
	 & $ 0.9728\pm 0.0035$ 
	 & $ 0.9728\pm 0.0035$ 
	 & $ 0.9747\pm 0.0035$ 
	 \\
$\tau_\mathrm{reio}$
	 & $ 0.0546\pm 0.0052$ 
	 & $ 0.0551\pm 0.0073$ 
	 & $ 0.0559^{+0.0069}_{-0.0077}$ 
	 & $ 0.0559\pm 0.0075$ 
	 & $ 0.0556\pm 0.0052$ 
	 \\
$\Omega_m$
	 & $ 0.3001\pm 0.0033$ 
	 & $ 0.2987\pm 0.0034$ 
	 & $ 0.2987\pm 0.0033$ 
	 & $ 0.2989\pm 0.0033$ 
	 & $ 0.2997\pm 0.0033$ 
	 \\
$\sigma_8$
	 & $ 0.8012\pm 0.0047$ 
	 & $ 0.7998\pm 0.0062$ 
	 & $ 0.7998\pm 0.0063$ 
	 & $ 0.8000\pm 0.0064$ 
	 & $ 0.7962^{+0.0059}_{-0.0053}$ 
	 \\
$S_8$
	 & $ 0.8014\pm 0.0076$ 
	 & $ 0.7981\pm 0.0085$ 
	 & $ 0.7981\pm 0.0084$ 
	 & $ 0.7985\pm 0.0085$ 
	 & $ 0.7958\pm 0.0083$ 
	 \\
\hline \hline 
$\Omega_{\rm DER}$
	 & 	
	 & $< 0.00622$ 
	 & $ 0.00262^{+0.00032}_{-0.0024}$ 
	 & 	
	 & 	
	 \\
$\log_{10}c_{\rm DER}$
	 & 	
	 & $ 6.92\pm 0.87$ 
	 & $> -0.106$ 
	 & 	
	 & 	
	 \\
$C_{\rm scf}$ $[M_{\text{pl}}\text{Mpc}^{-2}]$
	 & 	
	 & $< 3.63\times 10^{-7}$ 
	 & 	
	 & $< 4.00\times 10^{-8}$ 
	 & 	
	 \\
$\Omega_{\rm scf,k}$
	 & 	
	 & $< 1.32\times 10^{-4}$ 
	 & $< 5.94\times 10^{-6}$ 
	 & $< 0.00716$ 
	 & 	
	 \\
$w_0$
	 & 	
	 & 	
	 & 	
	 & 	
	 & $< -0.980$ 
	 \\
$w_a$
	 & 	
	 & 	
	 & 	
	 & 	
	 & $ 0.015^{+0.011}_{-0.024}$ 
	 \\

\hline \hline

$\chi^2$
	 & $ 1943.5\pm 6.6$ 
	 & $ 1944.3\pm 6.9$ 
	 & $ 1945.1\pm 6.7$ 
	 & $ 1945.5\pm 7.1$ 
	 & $ 1946.7\pm 7.1$ 
	 \\
$\chi^2_\mathrm{BAO\, DR12}$
	 & $ 3.52\pm 0.19$ 
	 & $ 3.51\pm 0.22$ 
	 & $ 3.50\pm 0.21$ 
	 & $ 3.50\pm 0.22$ 
	 & $ 3.44\pm 0.24$ 
	 \\
$\chi^2_\mathrm{BAO\ DR7\ MGS}$
	 & $ 2.08\pm 0.29$ 
	 & $ 2.09\pm 0.30$ 
	 & $ 2.09\pm 0.29$ 
	 & $ 2.09\pm 0.29$ 
	 & $ 2.09\pm 0.29$ 
	 \\
$\chi^2_\mathrm{SNe}$
	 & $ 67.7\pm 1.7$ 
	 & $ 68.1\pm 2.0$ 
	 & $ 68.2\pm 2.0$ 
	 & $ 68.3\pm 2.1$ 
	 & $ 69.1\pm 2.6$ 
	 \\
$\chi^2_\mathrm{Planck\ lowTT}$
	 & $ 21.78\pm 0.68$ 
	 & $ 21.96\pm 0.81$ 
	 & $ 21.88\pm 0.71$ 
	 & $ 21.88\pm 0.71$ 
	 & $ 21.59\pm 0.66$ 
	 \\
$\chi^2_\mathrm{Planck\ lowEE}$
	 & $ 396.40\pm 0.91$ 
	 & $ 396.9\pm 1.6$ 
	 & $ 397.0\pm 1.8$ 
	 & $ 397.0\pm 1.8$ 
	 & $ 396.5\pm 1.0$ 
	 \\
$\chi^2_\mathrm{CMB\ high-\ell}$
	 & $ 1452.0\pm 6.6$ 
	 & $ 1451.8\pm 6.7$ 
	 & $ 1452.4\pm 6.6$ 
	 & $ 1452.6\pm 7.0$ 
	 & $ 1453.9\pm 6.9$ 
	 \\
\hline 
    \end{tabular}
    \caption{
    Marginalised 1D $1\sigma$ posteriors and means are shown for various cosmologies described in Section~\ref{subsec:models}, constrained using forecasts produced with a fiducial \LCDM\ model as described in Section~\ref{subsec:data}.  
    Here $c_{\rm DER} = c_3$ in units of $[ 3^{-\frac{3}{4}}M_{\text{pl}}^{-\frac{3}{2}} \text{Mpc}^{\frac{1}{2}}]$ for mDER and $c_0$ in units of $[\text{Mpc}^{-1}]$ for tDER. 
    For one-sided bounds, we quote the 95\% upper or lower limit. 
    }
    \label{tab:results_future}
\end{table*}

\bibliographystyle{unsrt}
\bibliography{refs}

\begin{thebibliography}{100}

\bibitem{SupernovaSearchTeam:1998fmf}
Adam~G. Riess et~al.
\newblock {Observational evidence from supernovae for an accelerating universe
  and a cosmological constant}.
\newblock {\em Astron. J.}, 116:1009--1038, 1998.

\bibitem{SupernovaCosmologyProject:1998vns}
S.~Perlmutter et~al.
\newblock {Measurements of $\Omega$ and $\Lambda$ from 42 high redshift
  supernovae}.
\newblock {\em Astrophys. J.}, 517:565--586, 1999.

\bibitem{Scolnic:2017caz}
D.~M. Scolnic et~al.
\newblock {The Complete Light-curve Sample of Spectroscopically Confirmed SNe
  Ia from Pan-STARRS1 and Cosmological Constraints from the Combined Pantheon
  Sample}.
\newblock {\em Astrophys. J.}, 859(2):101, 2018.

\bibitem{Brout:2022vxf}
Dillon Brout et~al.
\newblock {The Pantheon+ Analysis: Cosmological Constraints}.
\newblock {\em Astrophys. J.}, 938(2):110, 2022.

\bibitem{Sherwin:2011gv}
Blake~D. Sherwin et~al.
\newblock {Evidence for dark energy from the cosmic microwave background alone
  using the Atacama Cosmology Telescope lensing measurements}.
\newblock {\em Phys. Rev. Lett.}, 107:021302, 2011.

\bibitem{Planck:2018vyg}
N.~Aghanim et~al.
\newblock {Planck 2018 results. VI. Cosmological parameters}.
\newblock {\em Astron. Astrophys.}, 641:A6, 2020.
\newblock [Erratum: Astron.Astrophys. 652, C4 (2021)].

\bibitem{Ross:2014qpa}
Ashley~J. Ross, Lado Samushia, Cullan Howlett, Will~J. Percival, Angela Burden,
  and Marc Manera.
\newblock {The clustering of the SDSS DR7 main Galaxy sample – I. A 4 per
  cent distance measure at $z = 0.15$}.
\newblock {\em Mon. Not. Roy. Astron. Soc.}, 449(1):835--847, 2015.

\bibitem{Alam:2016hwk}
Shadab Alam et~al.
\newblock {The clustering of galaxies in the completed SDSS-III Baryon
  Oscillation Spectroscopic Survey: cosmological analysis of the DR12 galaxy
  sample}.
\newblock {\em Mon. Not. Roy. Astron. Soc.}, 470(3):2617--2652, 2017.

\bibitem{Abbott:1984qf}
L.~F. Abbott.
\newblock {A Mechanism for Reducing the Value of the Cosmological Constant}.
\newblock {\em Phys. Lett. B}, 150:427--430, 1985.

\bibitem{PhysRevLett.52.1461}
T.~Banks.
\newblock Relaxation of the cosomological constant.
\newblock {\em Phys. Rev. Lett.}, 52:1461--1463, Apr 1984.

\bibitem{Alberte:2016izw}
Lasma Alberte, Paolo Creminelli, Andrei Khmelnitsky, David Pirtskhalava, and
  Enrico Trincherini.
\newblock {Relaxing the Cosmological Constant: a Proof of Concept}.
\newblock {\em JHEP}, 12:022, 2016.

\bibitem{PhysRevD.65.126003}
Paul~J. Steinhardt and Neil Turok.
\newblock Cosmic evolution in a cyclic universe.
\newblock {\em Phys. Rev. D}, 65:126003, May 2002.

\bibitem{Graham:2017hfr}
Peter~W. Graham, David~E. Kaplan, and Surjeet Rajendran.
\newblock {Born again universe}.
\newblock {\em Phys. Rev. D}, 97(4):044003, 2018.

\bibitem{Graham:2019bfu}
Peter~W. Graham, David~E. Kaplan, and Surjeet Rajendran.
\newblock {Relaxation of the Cosmological Constant}.
\newblock {\em Phys. Rev. D}, 100(1):015048, 2019.

\bibitem{Berghaus:2020ekh}
Kim~V. Berghaus, Peter~W. Graham, David~E. Kaplan, Guy~D. Moore, and Surjeet
  Rajendran.
\newblock {Dark energy radiation}.
\newblock {\em Phys. Rev. D}, 104(8):083520, 2021.

\bibitem{PhysRevD.37.3406}
Bharat Ratra and P.~J.~E. Peebles.
\newblock Cosmological consequences of a rolling homogeneous scalar field.
\newblock {\em Phys. Rev. D}, 37:3406--3427, Jun 1988.

\bibitem{Wetterich:1987fm}
C.~Wetterich.
\newblock {Cosmology and the Fate of Dilatation Symmetry}.
\newblock {\em Nucl. Phys. B}, 302:668--696, 1988.

\bibitem{Caldwell:1997ii}
R.~R. Caldwell, Rahul Dave, and Paul~J. Steinhardt.
\newblock {Cosmological imprint of an energy component with general equation of
  state}.
\newblock {\em Phys. Rev. Lett.}, 80:1582--1585, 1998.

\bibitem{osti_957435}
Daniel Green, Bart Horn, Phys.~Dept. /SLAC /Stanford~U., Leonardo Senatore,
  Phys. Dept. /Harvard-Smithsonian Ctr.~Astrophys. /Princeton, Inst. Advanced
  Study /Harvard~U., Eva Silverstein, and Phys.~Dept. /SLAC /Stanford~U.
\newblock Trapped inflation.
\newblock {\em Undecided}.

\bibitem{Anber:2009ua}
Mohamed~M. Anber and Lorenzo Sorbo.
\newblock {Naturally inflating on steep potentials through electromagnetic
  dissipation}.
\newblock {\em Phys. Rev. D}, 81:043534, 2010.

\bibitem{Fonseca:2019ypl}
Nayara Fonseca, Enrico Morgante, Ryosuke Sato, and G\'eraldine Servant.
\newblock {Axion fragmentation}.
\newblock {\em JHEP}, 04:010, 2020.

\bibitem{DallAgata:2019yrr}
Gianguido Dall'Agata, Sergio Gonz\'alez-Mart\'\i{}n, Alexandros Papageorgiou,
  and Marco Peloso.
\newblock {Warm dark energy}.
\newblock {\em JCAP}, 08:032, 2020.

\bibitem{Alexander:2022own}
Stephon Alexander, Heliudson Bernardo, and Michael~W. Toomey.
\newblock {Addressing the Hubble and $S_8$ Tensions with a Kinetically Mixed
  Dark Sector}.
\newblock 7 2022.

\bibitem{Bastero-Gil:2016qru}
Mar Bastero-Gil, Arjun Berera, Rudnei~O. Ramos, and Joao~G. Rosa.
\newblock {Warm Little Inflaton}.
\newblock {\em Phys. Rev. Lett.}, 117(15):151301, 2016.

\bibitem{Berghaus:2019whh}
Kim~V. Berghaus, Peter~W. Graham, and David~E. Kaplan.
\newblock {Minimal Warm Inflation}.
\newblock {\em JCAP}, 03:034, 2020.

\bibitem{Berera:1995ie}
Arjun Berera.
\newblock {Warm inflation}.
\newblock {\em Phys. Rev. Lett.}, 75:3218--3221, 1995.

\bibitem{Berera:1995wh}
Arjun Berera and Li-Zhi Fang.
\newblock {Thermally induced density perturbations in the inflation era}.
\newblock {\em Phys. Rev. Lett.}, 74:1912--1915, 1995.

\bibitem{Berera:1999ws}
Arjun Berera.
\newblock {Warm inflation at arbitrary adiabaticity: A Model, an existence
  proof for inflationary dynamics in quantum field theory}.
\newblock {\em Nucl. Phys.}, B585:666--714, 2000.

\bibitem{Berera:1998px}
Arjun Berera, Marcelo Gleiser, and Rudnei~O. Ramos.
\newblock {A First principles warm inflation model that solves the cosmological
  horizon / flatness problems}.
\newblock {\em Phys. Rev. Lett.}, 83:264--267, 1999.

\bibitem{Berera:2008ar}
Arjun Berera, Ian~G. Moss, and Rudnei~O. Ramos.
\newblock {Warm Inflation and its Microphysical Basis}.
\newblock {\em Rept. Prog. Phys.}, 72:026901, 2009.

\bibitem{Bastero-Gil:2010dgy}
Mar Bastero-Gil, Arjun Berera, and Rudnei~O. Ramos.
\newblock {Dissipation coefficients from scalar and fermion quantum field
  interactions}.
\newblock {\em JCAP}, 09:033, 2011.

\bibitem{Laine:2016hma}
Mikko Laine and Aleksi Vuorinen.
\newblock {\em {Basics of Thermal Field Theory}}, volume 925.
\newblock Springer, 2016.

\bibitem{Agrawal:2022yvu}
Prateek Agrawal, Kim~V. Berghaus, JiJi Fan, Anson Hook, Gustavo
  Marques-Tavares, and Tom Rudelius.
\newblock {Some open questions in axion theory}.
\newblock In {\em {2022 Snowmass Summer Study}}, 3 2022.

\bibitem{Moore:2010jd}
Guy~D. Moore and Marcus Tassler.
\newblock {The Sphaleron Rate in SU(N) Gauge Theory}.
\newblock {\em JHEP}, 02:105, 2011.

\bibitem{Chris_Graham_2009}
Chris Graham and Ian~G. Moss.
\newblock Density fluctuations from warm inflation.
\newblock {\em Journal of Cosmology and Astroparticle Physics}, 2009(07):013,
  jul 2009.

\bibitem{Bastero-Gil:2011rva}
Mar Bastero-Gil, Arjun Berera, and Rudnei~O. Ramos.
\newblock {Shear viscous effects on the primordial power spectrum from warm
  inflation}.
\newblock {\em JCAP}, 07:030, 2011.

\bibitem{Chevallier:2000qy}
Michel Chevallier and David Polarski.
\newblock {Accelerating universes with scaling dark matter}.
\newblock {\em Int. J. Mod. Phys. D}, 10:213--224, 2001.

\bibitem{Linder:2002et}
Eric~V. Linder.
\newblock {Exploring the expansion history of the universe}.
\newblock {\em Phys. Rev. Lett.}, 90:091301, 2003.

\bibitem{Fang:2008sn}
Wenjuan Fang, Wayne Hu, and Antony Lewis.
\newblock {Crossing the Phantom Divide with Parameterized Post-Friedmann Dark
  Energy}.
\newblock {\em Phys. Rev. D}, 78:087303, 2008.

\bibitem{Ballesteros:2010ks}
Guillermo Ballesteros and Julien Lesgourgues.
\newblock {Dark energy with non-adiabatic sound speed: initial conditions and
  detectability}.
\newblock {\em JCAP}, 10:014, 2010.

\bibitem{Lewis:2002ah}
Antony Lewis and Sarah Bridle.
\newblock {Cosmological parameters from CMB and other data: A Monte Carlo
  approach}.
\newblock {\em \prd}, 66:103511, 2002.

\bibitem{Lewis:2013hha}
Antony Lewis.
\newblock {Efficient sampling of fast and slow cosmological parameters}.
\newblock {\em Phys. Rev.}, D87(10):103529, 2013.

\bibitem{Torrado:2020dgo}
Jesus Torrado and Antony Lewis.
\newblock {Cobaya: Code for Bayesian Analysis of hierarchical physical models}.
\newblock {\em JCAP}, 05:057, 2021.

\bibitem{2019ascl.soft10019T}
Jes{\'u}s {Torrado} and Antony {Lewis}.
\newblock {Cobaya: Bayesian analysis in cosmology}.
\newblock Astrophysics Source Code Library, record ascl:1910.019, October 2019.

\bibitem{Blas:2011rf}
Diego Blas, Julien Lesgourgues, and Thomas Tram.
\newblock {The Cosmic Linear Anisotropy Solving System (CLASS) II:
  Approximation schemes}.
\newblock {\em JCAP}, 1107:034, 2011.

\bibitem{Lewis:1999bs}
Antony Lewis, Anthony Challinor, and Anthony Lasenby.
\newblock {Efficient computation of CMB anisotropies in closed FRW models}.
\newblock {\em Astrophys. J.}, 538:473--476, 2000.

\bibitem{Howlett:2012mh}
Cullan Howlett, Antony Lewis, Alex Hall, and Anthony Challinor.
\newblock {CMB power spectrum parameter degeneracies in the era of precision
  cosmology}.
\newblock {\em JCAP}, 1204:027, 2012.

\bibitem{Aghanim:2019ame}
N.~Aghanim et~al.
\newblock {Planck 2018 results. V. CMB power spectra and likelihoods}.
\newblock {\em Astron. Astrophys.}, 641:A5, 2020.

\bibitem{Hounsell:2017ejq}
R.~Hounsell et~al.
\newblock {Simulations of the WFIRST Supernova Survey and Forecasts of
  Cosmological Constraints}.
\newblock {\em Astrophys. J.}, 867(1):23, 2018.

\bibitem{Lewis:2019xzd}
Antony Lewis.
\newblock {GetDist: a Python package for analysing Monte Carlo samples}.
\newblock 2019.

\bibitem{DiValentino:2020vvd}
Eleonora Di~Valentino et~al.
\newblock {Cosmology Intertwined III: $f \sigma_8$ and $S_8$}.
\newblock {\em Astropart. Phys.}, 131:102604, 2021.

\bibitem{Abdalla:2022yfr}
Elcio Abdalla et~al.
\newblock {Cosmology intertwined: A review of the particle physics,
  astrophysics, and cosmology associated with the cosmological tensions and
  anomalies}.
\newblock {\em JHEAp}, 34:49--211, 2022.

\bibitem{Wolf:2023uno}
William~J. Wolf and Pedro~G. Ferreira.
\newblock {Underdetermination of dark energy}.
\newblock {\em Phys. Rev. D}, 108(10):103519, 2023.

\bibitem{Secco:2022kqg}
Lucas~F. Secco, Tanvi Karwal, Wayne Hu, and Elisabeth Krause.
\newblock {Role of the Hubble scale in the weak lensing versus CMB tension}.
\newblock {\em Phys. Rev. D}, 107(8):083532, 2023.

\bibitem{Bakhti:2013ora}
Pouya Bakhti and Yasaman Farzan.
\newblock {Constraining Super-light Sterile Neutrino Scenario by JUNO and
  RENO-50}.
\newblock {\em JHEP}, 10:200, 2013.

\bibitem{Chen:2022zts}
Zikang Chen, Jiajun Liao, Jiajie Ling, and Baobiao Yue.
\newblock {Constraining super-light sterile neutrinos at Borexino and KamLAND}.
\newblock {\em JHEP}, 09:004, 2022.

\bibitem{PTOLEMY:2018jst}
E.~Baracchini et~al.
\newblock {PTOLEMY: A Proposal for Thermal Relic Detection of Massive Neutrinos
  and Directional Detection of MeV Dark Matter}.
\newblock 8 2018.

\bibitem{PTOLEMY:2019hkd}
M.~G. Betti et~al.
\newblock {Neutrino physics with the PTOLEMY project: active neutrino
  properties and the light sterile case}.
\newblock {\em JCAP}, 07:047, 2019.

\bibitem{Zhang:2017ljh}
Jue Zhang and Xin Zhang.
\newblock {Gravitational clustering of cosmic relic neutrinos in the Milky
  Way}.
\newblock {\em Nature Commun.}, 9:1833, 2018.

\bibitem{Long:2014zva}
Andrew~J. Long, Cecilia Lunardini, and Eray Sabancilar.
\newblock {Detecting non-relativistic cosmic neutrinos by capture on tritium:
  phenomenology and physics potential}.
\newblock {\em JCAP}, 08:038, 2014.

\bibitem{Cheipesh:2021fmg}
Yevheniia Cheipesh, Vadim Cheianov, and Alexey Boyarsky.
\newblock {Navigating the pitfalls of relic neutrino detection}.
\newblock {\em Phys. Rev. D}, 104(11):116004, 2021.

\bibitem{PTOLEMY:2022ldz}
A.~Apponi et~al.
\newblock {Heisenberg\textquoteright{}s uncertainty principle in the PTOLEMY
  project: A theory update}.
\newblock {\em Phys. Rev. D}, 106(5):053002, 2022.

\bibitem{Cavoto:2022xwo}
Gianluca Cavoto, Angelo Esposito, Guglielmo Papiri, and Antonio~Davide Polosa.
\newblock {Relativistic corrections to polarized-tritium \ensuremath{\beta}
  decay}.
\newblock {\em Phys. Rev. C}, 107(6):064603, 2023.

\bibitem{Tullynew}
personal communication~with Christopher~Tully.
\newblock {March 24th, 2023}.

\bibitem{CAST:2017uph}
V.~Anastassopoulos et~al.
\newblock {New CAST Limit on the Axion-Photon Interaction}.
\newblock {\em Nature Phys.}, 13:584--590, 2017.

\bibitem{Ayala:2014pea}
Adrian Ayala, Inma Dom\'\i{}nguez, Maurizio Giannotti, Alessandro Mirizzi, and
  Oscar Straniero.
\newblock {Revisiting the bound on axion-photon coupling from Globular
  Clusters}.
\newblock {\em Phys. Rev. Lett.}, 113(19):191302, 2014.

\bibitem{Carenza:2020zil}
Pierluca Carenza, Oscar Straniero, Babette D\"obrich, Maurizio Giannotti,
  Giuseppe Lucente, and Alessandro Mirizzi.
\newblock {Constraints on the coupling with photons of heavy
  axion-like-particles from Globular Clusters}.
\newblock {\em Phys. Lett. B}, 809:135709, 2020.

\bibitem{AxionLimits}
Ciaran O'Hare.
\newblock cajohare/axionlimits: Axionlimits.
\newblock \url{https://cajohare.github.io/AxionLimits/}, July 2020.

\bibitem{Caldwell:2016dcw}
Allen Caldwell, Gia Dvali, B\'ela Majorovits, Alexander Millar, Georg Raffelt,
  Javier Redondo, Olaf Reimann, Frank Simon, and Frank Steffen.
\newblock {Dielectric Haloscopes: A New Way to Detect Axion Dark Matter}.
\newblock {\em Phys. Rev. Lett.}, 118(9):091801, 2017.

\bibitem{Lawson:2019brd}
Matthew Lawson, Alexander~J. Millar, Matteo Pancaldi, Edoardo Vitagliano, and
  Frank Wilczek.
\newblock {Tunable axion plasma haloscopes}.
\newblock {\em Phys. Rev. Lett.}, 123(14):141802, 2019.

\bibitem{Alesini:2020vny}
D.~Alesini et~al.
\newblock {Search for invisible axion dark matter of mass m$_a=43~\mu$eV with
  the QUAX--$a\gamma$ experiment}.
\newblock {\em Phys. Rev. D}, 103(10):102004, 2021.

\bibitem{BREAD:2021tpx}
Jesse Liu et~al.
\newblock {Broadband Solenoidal Haloscope for Terahertz Axion Detection}.
\newblock {\em Phys. Rev. Lett.}, 128(13):131801, 2022.

\bibitem{PhysRevD.102.043003}
Graciela~B. Gelmini, Alexander~J. Millar, Volodymyr Takhistov, and Edoardo
  Vitagliano.
\newblock Probing dark photons with plasma haloscopes.
\newblock {\em Phys. Rev. D}, 102:043003, Aug 2020.

\bibitem{PhysRevLett.123.141802}
Matthew Lawson, Alexander~J. Millar, Matteo Pancaldi, Edoardo Vitagliano, and
  Frank Wilczek.
\newblock Tunable axion plasma haloscopes.
\newblock {\em Phys. Rev. Lett.}, 123:141802, Oct 2019.

\bibitem{DMRadio:2022jfv}
L.~Brouwer et~al.
\newblock {Proposal for a definitive search for GUT-scale QCD axions}.
\newblock {\em Phys. Rev. D}, 106(11):112003, 2022.

\bibitem{Aja:2022csb}
Beatriz Aja et~al.
\newblock {The Canfranc Axion Detection Experiment (CADEx): search for axions
  at 90 GHz with Kinetic Inductance Detectors}.
\newblock {\em JCAP}, 11:044, 2022.

\bibitem{Ahyoune:2023gfw}
S.~Ahyoune et~al.
\newblock {A proposal for a low-frequency axion search in the 1-2 $\mu$eV range
  and below with the BabyIAXO magnet}.
\newblock 6 2023.

\bibitem{DeMiguel:2023nmz}
Javier De~Miguel and Juan~F. Hern\'andez-Cabrera.
\newblock {Discovery prospects with the Dark-photons \& Axion-Like particles
  Interferometer:part I}.
\newblock 3 2023.

\bibitem{ADMX:2001nej}
Stephen~J. Asztalos et~al.
\newblock {Experimental constraints on the axion dark matter halo density}.
\newblock {\em Astrophys. J. Lett.}, 571:L27--L30, 2002.

\bibitem{ADMX:2018gho}
N.~Du et~al.
\newblock {A Search for Invisible Axion Dark Matter with the Axion Dark Matter
  Experiment}.
\newblock {\em Phys. Rev. Lett.}, 120(15):151301, 2018.

\bibitem{ADMX:2019uok}
T.~Braine et~al.
\newblock {Extended Search for the Invisible Axion with the Axion Dark Matter
  Experiment}.
\newblock {\em Phys. Rev. Lett.}, 124(10):101303, 2020.

\bibitem{ADMX:2023rsk}
T.~Nitta et~al.
\newblock {Search for the Cosmic Axion Background with ADMX}.
\newblock 3 2023.

\bibitem{HAYSTAC:2018rwy}
L.~Zhong et~al.
\newblock {Results from phase 1 of the HAYSTAC microwave cavity axion
  experiment}.
\newblock {\em Phys. Rev. D}, 97(9):092001, 2018.

\bibitem{HAYSTAC:2020kwv}
K.~M. Backes et~al.
\newblock {A quantum-enhanced search for dark matter axions}.
\newblock {\em Nature}, 590(7845):238--242, 2021.

\bibitem{HAYSTAC:2023cam}
M.~J. Jewell et~al.
\newblock {New results from HAYSTAC\textquoteright{}s phase II operation with a
  squeezed state receiver}.
\newblock {\em Phys. Rev. D}, 107(7):072007, 2023.

\bibitem{Salemi:2021gck}
Chiara~P. Salemi et~al.
\newblock {Search for Low-Mass Axion Dark Matter with ABRACADABRA-10~cm}.
\newblock {\em Phys. Rev. Lett.}, 127(8):081801, 2021.

\bibitem{Dror:2021nyr}
Jeff~A. Dror, Hitoshi Murayama, and Nicholas~L. Rodd.
\newblock {Cosmic axion background}.
\newblock {\em Phys. Rev. D}, 103(11):115004, 2021.
\newblock [Erratum: Phys.Rev.D 106, 119902 (2022)].

\bibitem{Chounew}
Aaron~S. Chou and Pierre~M. Echternach.
\newblock {Manuscript in preparation}.
\newblock {2024}.

\bibitem{Caputo:2021eaa}
Andrea Caputo, Alexander~J. Millar, Ciaran A.~J. O'Hare, and Edoardo
  Vitagliano.
\newblock {Dark photon limits: A handbook}.
\newblock {\em Phys. Rev. D}, 104(9):095029, 2021.

\bibitem{PhysRevLett.61.2285}
D.~F. Bartlett and Stefan L\"ogl.
\newblock Limits on an electromagnetic fifth force.
\newblock {\em Phys. Rev. Lett.}, 61:2285--2287, Nov 1988.

\bibitem{Kroff:2020zhp}
D.~Kroff and P.~C. Malta.
\newblock {Constraining hidden photons via atomic force microscope measurements
  and the Plimpton-Lawton experiment}.
\newblock {\em Phys. Rev. D}, 102(9):095015, 2020.

\bibitem{Ehret:2010mh}
Klaus Ehret et~al.
\newblock {New ALPS Results on Hidden-Sector Lightweights}.
\newblock {\em Phys. Lett. B}, 689:149--155, 2010.

\bibitem{Romanenko:2023irv}
A.~Romanenko et~al.
\newblock {Search for Dark Photons with Superconducting Radio Frequency
  Cavities}.
\newblock {\em Phys. Rev. Lett.}, 130(26):261801, 2023.

\bibitem{Fixsen:1996nj}
D.~J. Fixsen, E.~S. Cheng, J.~M. Gales, John~C. Mather, R.~A. Shafer, and E.~L.
  Wright.
\newblock {The Cosmic Microwave Background spectrum from the full COBE FIRAS
  data set}.
\newblock {\em Astrophys. J.}, 473:576, 1996.

\bibitem{Mather:1998gm}
John~C. Mather, D.~J. Fixsen, R.~A. Shafer, C.~Mosier, and D.~T. Wilkinson.
\newblock {Calibrator design for the COBE far infrared absolute
  spectrophotometer (FIRAS)}.
\newblock {\em Astrophys. J.}, 512:511--520, 1999.

\bibitem{Mirizzi:2009iz}
Alessandro Mirizzi, Javier Redondo, and Gunter Sigl.
\newblock {Microwave Background Constraints on Mixing of Photons with Hidden
  Photons}.
\newblock {\em JCAP}, 03:026, 2009.

\bibitem{McDermott:2019lch}
Samuel~D. McDermott and Samuel~J. Witte.
\newblock {Cosmological evolution of light dark photon dark matter}.
\newblock {\em Phys. Rev. D}, 101(6):063030, 2020.

\bibitem{Caputo:2020bdy}
Andrea Caputo, Hongwan Liu, Siddharth Mishra-Sharma, and Joshua~T. Ruderman.
\newblock {Dark Photon Oscillations in Our Inhomogeneous Universe}.
\newblock {\em Phys. Rev. Lett.}, 125(22):221303, 2020.

\bibitem{Caputo:2020rnx}
Andrea Caputo, Hongwan Liu, Siddharth Mishra-Sharma, and Joshua~T. Ruderman.
\newblock {Modeling Dark Photon Oscillations in Our Inhomogeneous Universe}.
\newblock {\em Phys. Rev. D}, 102(10):103533, 2020.

\bibitem{Schwarz:2015lqa}
Matthias Schwarz, Ernst-Axel Knabbe, Axel Lindner, Javier Redondo, Andreas
  Ringwald, Magnus Schneide, Jaroslaw Susol, and G\"unter Wiedemann.
\newblock {Results from the Solar Hidden Photon Search (SHIPS)}.
\newblock {\em JCAP}, 08:011, 2015.

\bibitem{An:2020bxd}
Haipeng An, Maxim Pospelov, Josef Pradler, and Adam Ritz.
\newblock {New limits on dark photons from solar emission and keV scale dark
  matter}.
\newblock {\em Phys. Rev. D}, 102:115022, 2020.

\bibitem{XENON:2021qze}
E.~Aprile et~al.
\newblock {Emission of single and few electrons in XENON1T and limits on light
  dark matter}.
\newblock {\em Phys. Rev. D}, 106(2):022001, 2022.

\bibitem{Li:2023vpv}
Shao-Ping Li and Xun-Jie Xu.
\newblock {Production rates of dark photons and $Z'$ in the Sun and stellar
  cooling bounds}.
\newblock 4 2023.

\bibitem{A_Kogut_2011}
A.~Kogut, D.J. Fixsen, D.T. Chuss, J.~Dotson, E.~Dwek, M.~Halpern, G.F.
  Hinshaw, S.M. Meyer, S.H. Moseley, M.D. Seiffert, D.N. Spergel, and E.J.
  Wollack.
\newblock The primordial inflation explorer ({PIXIE}): a nulling polarimeter
  for cosmic microwave background observations.
\newblock {\em Journal of Cosmology and Astroparticle Physics},
  2011(07):025--025, jul 2011.

\bibitem{Kogut_2020}
A.~Kogut and D.J. Fixsen.
\newblock Calibration method and uncertainty for the primordial inflation
  explorer ({PIXIE}).
\newblock {\em Journal of Cosmology and Astroparticle Physics},
  2020(05):041--041, may 2020.

\bibitem{Pirvu:2023lch}
Dalila P\^\i{}rvu, Junwu Huang, and Matthew~C. Johnson.
\newblock {Patchy Screening of the CMB from Dark Photons}.
\newblock 7 2023.

\bibitem{Godfrey:2021tvs}
Benjamin Godfrey et~al.
\newblock {Search for dark photon dark matter: Dark E field radio pilot
  experiment}.
\newblock {\em Phys. Rev. D}, 104(1):012013, 2021.

\bibitem{Beaufort:2023qpd}
C.~Beaufort, M.~Bastero-Gil, A.~Catalano, D-S. Erfani-Harami, O.~Guillaudin,
  D.~Santos, S.~Savorgnano, and F.~Vezzu.
\newblock {Directional detection of meV dark photons with Dandelion}.
\newblock 10 2023.

\bibitem{2021JATIS...7a1003E}
Pierre~M. {Echternach}, Andrew~D. {Beyer}, and Charles~M. {Bradford}.
\newblock {Large array of low-frequency readout quantum capacitance detectors}.
\newblock {\em Journal of Astronomical Telescopes, Instruments, and Systems},
  7:011003, January 2021.

\bibitem{Stefanazzi:2021otz}
Leandro Stefanazzi et~al.
\newblock {The QICK (Quantum Instrumentation Control Kit): Readout and control
  for qubits and detectors}.
\newblock {\em Rev. Sci. Instrum.}, 93(4):044709, 2022.

\end{thebibliography}

\end{document}